\documentclass[11pt,english,paper]{revtex4}
\usepackage[T1]{fontenc}
\usepackage[latin9]{inputenc}
\usepackage{subfigure}
\usepackage{float}
\usepackage{amsmath}
\usepackage{color}
\usepackage{graphicx}
\usepackage{amssymb}



\usepackage{babel}

\begin{document}

\title{Unified theory of the semi-collisional tearing mode and internal
kink mode in a hot tokamak: implications for sawtooth modelling}

\author{J. W. Connor$^{1,2,3}$, R. J. Hastie$^{1,3}$ and A. Zocco$^{1,3}$}

\affiliation{$^{1}$Euratom/CCFE Fusion Association, Culham Science Centre, Abingdon,
Oxon, UK, OX14 3DB\\$^{2}$Imperial College of Science Technology
and Medicine, London SW7 2BZ, \\ $^{3}$Rudolf Peierls Centre for
Theoretical Physics, 1 Keble Road, Oxford, UK, OX1 3NP}

\begin{abstract}
In large hot tokamaks like JET, the width of the reconnecting layer
for resistive modes is determined by semi-collisional electron dynamics
and is much less than the ion Larmor radius. Firstly a dispersion
relation valid in this regime is derived which provides a unified
description of drift-tearing modes, kinetic Alfv\'en waves and the
internal kink mode at low beta. Tearing mode stability is investigated
analytically recovering the stabilising ion orbit effect, obtained
previously by Cowley et al. (Phys. Fluids \textbf{29 }3230 1986),
which implies large values of the tearing mode stability parameter
$\Delta^{\prime}$ are required for instability. Secondly, at high
beta it is shown that the tearing mode interacts with the kinetic
Alfv\'en wave and that there is an absolute stabilisation for all
$\Delta^{\prime}$ due to the shielding effects of the electron temperature
gradients, extending the result of Drake et. al (Phys. Fluids \textbf{26}
2509 1983) to large ion orbits. The nature of the transition between
these two limits at finite values of beta is then elucidated. The
low beta formalism is also relevant to the $m=n=1$ tearing mode and
the dissipative internal kink mode, thus extending the work of Pegoraro
et al. (Phys. Fluids B \textbf{1 }364 1989) to a more realistic electron
model incorporating temperature perturbations, but then the smallness
of the dissipative internal kink mode frequency is exploited to obtain
a new dispersion relation valid at arbitrary beta. A diagram describing
the stability of both the tearing mode and dissipative internal kink
mode, in the space of $\Delta^{\prime}$ and beta, is obtained. The
trajectory of the evolution of the current profile during a sawtooth
period can be plotted in this diagram, providing a model for the triggering
of a sawtooth crash.
\end{abstract}
\maketitle
\newpage

\section{Introduction}

Magnetised plasmas are potentially subject to electromagnetic tearing
mode instabilities. In cylindrical (or slab) geometry these instabilities
are driven by the potential energy associated with plasma currents
flowing along the magnetic field, characterised by a quantity $\Delta^{\prime}$
\citep{FKR}, and occur when magnetic reconnection due to dissipative
effects can take place in a narrow region around a ``resonant surface'',
i.e. where $k_{\parallel},$ the wave number of the tearing mode along
the magnetic field direction, is zero. Such surfaces can occur in
the presence of magnetic shear. The stability of this mode is given
by a dispersion relation $\Delta^{\prime}=\Delta^{\prime}_L(\omega),$
where $\Delta^{\prime}_L(\omega)$ is a function that depends on the plasma response at the resonant layer
and $\omega$ is the mode frequency. This instability is of importance
in both laboratory plasmas, such as those associated with magnetic
confinement fusion, and astrophysical plasmas.

Although many reconnection events involve nonlinear phenomena, two
situations in which linear theory can be important are sawteeth and
mode locking phenomena in tokamaks. In present day tokamaks such
as JET and the international burning plasma device ITER, currently
under construction in France, the $m=1,$ $n=1,$ tearing (or {}``reconnecting'')
mode, and related internal kink mode, where $m$ and $n$ are the
poloidal and toroidal mode numbers, are believed to trigger the sawtooth
instability, a periodic disturbance of the core of the toroidal plasma
column that occurs when the safety factor, $q$, falls below unity
and degrades performance. The rapid sawtooth event that is observed
in hot tokamaks \citep{Jim} could be associated with a high threshold
in $\Delta^{\prime}$ {[}sometimes replaced by $\lambda_{H}=-\pi/\Delta^{\prime}r_{s}$,
where $\lambda_{H}$ is proportional to the growth rate of the ideal
MHD internal kink mode \citep{coppi-res} and $q(r_{s})=1$], followed
by some nonlinear destabilisation mechanism that suddenly releases
the available magnetic energy. For example, the collisional Braginskii
two-fluid model predicts a stable window in plasma pressure between
the drift-tearing and the internal kink modes \citep{10.1063/1.859828},
so that the sawtooth crash may be associated with crossing the higher
pressure instability boundary. The Porcelli et al. model \citep{0741-3335-38-12-010}
employs a sawtooth trigger criterion given by the conditions for the
onset of growth of the resistive internal kink mode \citep{0741-3335-28-4-003,pegoraro:364},
but it involves some kinetic effects (a large Larmor radius model
for the ions and a rapid electron transport along the magnetic field).
The resulting criterion can be expressed in terms of a critical value
of the magnetic shear, $\hat{s}=rd(\ln q)/dr,$ at the rational $q=1$
surface. Such qualitative models are often
used for describing sawtooth behaviour in ITER \citep{0029-5515-33-3-I01}.
Determining the critical value of $\Delta^{\prime}$ for the instability
of these modes in such plasmas, as caculated from a relevant theory,
should help to model and control the sawtooth activity in these devices.

For mode locking phenomena, one needs to calculate the low level of
magnetic perturbation induced by an external coil at a mode resonant surface that produces an
electromagnetic torque sufficient to overcome viscous plasma forces \citep{10.1063/1.2178167,PhysRevLett.99.065001}.
This torque can be calculated in terms of the layer quantity $\Delta^{\prime}_L(\omega),$
but evaluated at a frequency $\omega$ representing the mismatch between the
$\mathbf E \times \mathbf B$ and diamagnetic rotation of the plasma at the layer and the
static external coil, rather than the natural frequency of a tearing mode.
This information is very important for the resonant magnetic perturbations
being considered for ELM control on ITER$\,$ \citep{PhysRevLett.92.235003}.

Initially, tearing modes were studied theoretically using the equations
of resistive MHD in cylindrical geometry, when instability occurs
if $\Delta^{\prime}>0$ \citep{FKR}, although a later toroidal calculation
obtained $\Delta^{\prime}>\Delta_{crit}^{\prime},$ where $\Delta_{crit}^{\prime}$
is a critical value related to the favourable average curvature \citep{10.1063/1.861224}.
However, for hotter plasmas various kinetic effects become important.
The inclusion of diamagnetic effects \citep{bussacsaw} leads to a
reduction of the growth rate and imposes a rotation with frequency
$\omega\sim\omega_{*e}=k_{y}T_{e}/(eBL_{n}),$ where $k_{y}$ is the
wave-number in the magnetic surface but perpendicular to the magnetic
field (in a tokamak $k_{y}=m/r$ ), $T_{e}$ is the electron temperature,
$e$ is the electron charge, $B$ the magnetic field strength and
$L_{n}=[d(\ln n_{e})/dr]^{-1}$ is the density scale-length. In this
case the mode is identified as a drift-tearing mode.

In contemporary fusion devices the {}``semicollisional'' regime
is relevant; this is a situation in which the electron collisional
transport along the perturbed magnetic field becomes comparable with
the mode frequency, i.e. $\omega\sim k_{\parallel}^{2}v_{the}^{2}/\nu_{e},$
where $v_{the}=(2T_{e}/m_{e})^{1/2}$ is the electron thermal speed, $m_{e}$ being the electron mass, and $\nu_{e}$ is the electron
collision frequency. Since $k_{\parallel}=k_{y}x/L_{s},$ where $L_{s}$
is the magnetic shear length (in a tokamak $L_s=R q /\hat s $ with $R$ the major radius), and $x$ the distance from the resonant
surface, this defines a semi-collisional width $\delta_{0}=L_{s}(\omega_{*e}\nu_{e})^{1/2}/(k_{y}v_{the})$\citep{ruthfurth,drake:1341,10.1063/1.863576,coppi-collisionless}.
When $\delta_{0}\gg\rho_{i},$ where $\rho_{i}=(2m_{i}T_{i})^{1/2}/eB$
is the ion Larmor radius, $m_{i}$ and $T_{i}$ being the ion mass
and temperature, respectively, a small Larmor radius approximation
can be employed for the ions, but the opposite limit can be more relevant.

Three key parameters govern such calculations: $\hat{\beta}=\beta_{e}(L_{s}/L_{n})^{2}/2,$
where $\beta_{e}=2\mu_{0}n_{e}T_{e}/B^{2}$ is the normalised electron
pressure, a collisionality parameter $C=0.51(\nu_{e}/\omega_{*e})(m_{e}/m_{i})(L_{s}/L_{n})^{2},$
and $\Delta^{\prime}.$ To see the significance of $\hat{\beta}$
we note that the collisional resistive layer width, $\delta_{\eta},$
i.e. the width of the current channel at the resonant surface, can
be estimated by balancing the mode frequency against the resistive
diffusion across this width: $\omega\sim\eta_{\parallel}/\delta_{\eta}^{2},$
where $\eta_{\parallel}$ is the resistivity parallel to the magnetic
field. For drift-tearing modes with $\omega\sim\omega_{*e}$ we find
that the ratio of resistive and semi-collisional widths is given by $(\delta_{\eta}/\delta_{0})^{2}\sim2/\hat{\beta},$ so that $\hat{\beta}$
is the key parameter controlling semi-collisional effects: for $\hat{\beta}\gg1$
the resistive layer width is much narrower than the semi-collisional
one so the semi-collisional effects dominate.

Regarding the parameter $C,$ we can write $C=0.51(\delta_{0}/\rho_{s})^{2},$
where $\rho_{s}=\tau^{1/2}\rho_{i}$ with $\tau=T_{e}/T_{i}$ (i.e.
$\rho_{s}$ is the ion Larmor radius based on $T_{e}$), so it measures
the ratio of the semi-collisional width to the ion Larmor radius.
Of course, a cold ion model, i.e. $\tau\rightarrow\infty,$ decouples
the parameter $C$ from the ratio of the ion Larmor radius to the
semi-collisional layer width and it is then unnecessary to account
for finite ion orbit effects. Drake et al. \citep{drake:2509} considered
such a model; in the limit $C\ll1$ they found that the tearing mode
is stable until $\Delta^{\prime}$ reaches a critical value: $\Delta_{crit}^{\prime}\sim\rho_{s}^{-1}\hat{\beta}$
for low values of $\hat{\beta}.$ For $\hat{\beta}\gg1,$ they found
that the equilibrium radial electron temperature gradients shield
the resonant surface from the magnetic perturbation and the tearing
mode is stable for all $\Delta^{\prime}.$ The cold ion model with
$\rho_{s}>\delta_{0}$ has, since then, been revisited by a number of authors using several, more or less ad-hoc, electron models. 
Thus Grasso et al. \cite{grassoop1ref} considered a low $\beta,$ three field model including
the effects of ion viscosity, particle diffusion and finite electron compressibility;
while growth rates were affected by electron diamagnetism, the instability threshold was found to be
unchanged. This model has been recently re-derived by Fitzpatrick \cite{fitzref} (in the absence of density and temperature gradients), and applied to collisionless and semi-collisional situations. In a subsequent paper
Grasso et al. \cite{grassoop2ref} extended their work to a four field model including ion motion along the magnetic field,
demonstrating the stabilising influence of ion acoustic waves when $\beta_e>\Delta^{\prime}\rho_s(L_n/L_s)^{1/2},$
discovered earlier by Bussac et al.\cite{Bussacref}. None of these models, however, can reproduce
the semi-collisional electron conductivity calculated in Ref. \cite{drake:2509}
because the authors ignored electron thermal effects. However, thermal effects including anomalous electron cross-field transport have been investigated numerically in Ref. \cite{nishimuraref}.  
The effect of toroidal geometry has also been studied \cite{hahmchen1,yuref}.

For a high temperature tokamak, however, the condition $\rho_{i}\ll\delta_{0}$
is only valid at very low values of magnetic shear $\hat{s},$ a situation
that might nevertheless arise at the $q=1$ surface where the $m=n=1$ tearing
mode is resonant. When the width of the reconnection layer becomes narrower than the
ion Larmor radius a fully gyro-kinetic description of the ions is
necessary, as noted by Antonsen and Coppi \citep{antonsen-coppi},
Hahm and Chen \citep{tshamthesis,hahmchen2}, Cowley et al. \citep{cowley:3230},
Pegoraro and Schep \citep{pegoraro:478}, and Pegoraro, Porcelli and
Schep \citep{pegoraro:364}. As a result of the non-local character
of the finite ion orbits, the problem involves the solution of a set
of integro-differential equations. Cowley et al. \citep{cowley:3230}
considered both collisionless and semi-collisional electron models.
In the collisionless case they recovered the strong stabilising effect
from the ion Larmor orbits found by Antonsen and Coppi \citep{antonsen-coppi},
but also found a similar effect in the semi-collisional case. This
is due to the non-local ion response to electrostatic perturbations.
Mathematically, it is essential to keep the small corrections arising
from the $(k\rho_{i})^{-1}$ tail at high $k\rho_{i}$ in $F(k\rho_{i}),$
the Fourier transform of the ion response with respect to distance
from the resonant surface. Calculations using a Pad\'e approximation
to $F(k\rho_{i}),$ as in part of Refs. \citep{pegoraro:478} and
\citep{pegoraro:364}, would fail to capture the effect of this tail.
The calculation by Cowley et al. \citep{cowley:3230}, which was restricted
to low $\hat{\beta},$ found the stabilising effect from large ion
orbits was characterised by $\Delta_{crit}^{\prime}\sim\rho_{i}^{-1}\eta_{e}^{2}\hat{\beta}\ln(\rho_{i}/\delta_{0}),$
where $\eta_{e}=d(\ln T_{e})/d(\ln n_{e}).$ The semi-collisional
calculations in Refs. \citep{pegoraro:364}, \citep{hahmchen2},
\citep{tshamthesis} and \citep{pegoraro:478}, were restricted to
the limits of negligible and infinite parallel heat conduction, missing
important thermal effects. In Ref. \citep{pegoraro:364} the large
ion orbit theory was applied to the stability of the ideal and dissipative
internal kink mode in a tokamak, which resulted in higher growth rates
than in simpler fluid models \citep{ABC}.

In this paper we revisit the semi-collisional situation in the $\rho_{i}\gg\delta_{0}$
limit addressed in Ref. \citep{cowley:3230}, but by using a different
mathematical approach (based on Fourier transforming the governing
integro-differential equations) extend it to finite $\hat{\beta}$
so that we can follow the transition from the stabilising effect on
tearing modes of finite ion orbits at low $\hat{\beta}$ to the stabilising
effect of the electron temperature gradient at high $\hat{\beta.}$
The problem can then be reduced to the solution of a fourth order
system of differential equations. We take advantage of the separation
of scales between the ion Larmor radius and the semi-collisional layer
width when $\delta_{0}/\rho_{i}\ll1,$ i.e. $C\ll1,$ to reduce this
system to the consideration of two separate regions: one the {}``ion
region'', where the effects of the finite ion Larmor radius appear
and another, the {}``electron region'', where the semi-collisional
effects on the electrons are dominant. The long wavelength limit of
the solution in the ion region must be matched to an ideal MHD solution
characterised by $\Delta^{\prime}.$ At short wavelengths the ion
region solution must be matched to the solution in the electron region,
leading to a dispersion relation.

The paper is organised as follows. In Section \ref{sec:formulation}
we formulate the tearing mode stability problem and the separation
into the ion and electron regions. In Section \ref{sec:lowbetaJIM}
we obtain a unified analytic dispersion relation for the coupled drift-tearing
mode and the FLR modified internal kink mode in the limit $\hat{\beta}\ll1.$
When $\Delta^{\prime}>0$ this dispersion relation exhibits coupling
between the drift-tearing branch and kinetic Alfv\'en waves and we
analyse the resulting stability of these modes as a function of the
parameter $\Delta^{\prime}\rho_{i}\hat{\beta}.$ In Section \ref{sec:highbeta}
we obtain an analytic solution in the opposite limit, $\hat{\beta}\gg1.$
The transition from instability at low $\hat{\beta}$ to stability
at high $\hat{\beta}$ is explored in Section \ref{sec:finitebeta},
using a finite $\hat{\beta}$ analysis, but one which is strictly
valid only for a unique value of $\eta_{e}.$ These results on tearing mode 
stability are summarised in Section \ref{sec:stabdiagr}.
Section \ref{sec:The-Internal-Kink}
analyses the stability of the internal kink mode, emphasizing the
dissipative mode which is relevant when $\Delta^{\prime}>0.$ As well
as applying the unified low $\hat{\beta}$ dispersion relation to
this mode, we also exploit the smallness of its frequency to obtain
a dispersion relation valid for finite $\hat{\beta}$ in the Appendix which is also studied in Section \ref{sec:The-Internal-Kink}.
In Section \ref{sec:Sawtooth-Modelling} we address the implications
of the stability analysis for sawtooth modelling. Finally the results
are discussed and summarised in Sec. \ref{sec:Conclusions}.

\section{Semi-Collisional Stability Equations\label{sec:formulation}}

We consider a plasma immersed in a sheared magnetic field in slab geometry, $\mathbf B_0=B_0(\mathbf z +x/L_s \mathbf y)$,
which serves as an approximation to a cylindrical plasma provided we ignore the effects
of drifts due to the cylindrical magnetic curvature. Thus $B_0$ is the strength of the guide field and $L_s$ the shear length.
A gyro-kinetic description of the ions allows us to model finite ion Larmor orbits, since we consider these to be large
in comparison to the width of the reconnecting layer where the electron current flows,
$\rho_i>\delta_0;$ however we do not include the effects of ion motion along the magnetic field.
Since we do not attempt to describe toroidal geometry, the effects of finite ion orbits
due to inhomogeneous magnetic fields and their neoclassical consequences cannot be addressed.
The electrons are described by a semi-collisional fluid model based on an Ohm's law
with a thermal force and an electron thermal equation in which parallel electron thermal transport
competes with the mode frequency. Again, the limitation to slab geometry precludes neoclassical 
effects arising from trapped particles. These ion and electron models allow us to calculate
the perturbed charge densities and current produced by perturbations in the electrostatic 
and vector potentials and hence, by introducing these in the quasineutrality condition and Amp\'ere's law,
we arrive at a coupled set of eigenvalue equation. 

Such a model has been provided by Cowley et al. \citep{cowley:3230} and we reproduce it here: $\rho_{i}\geq\delta$\begin{equation}\label{eq:eq1}
-\frac{x}{\delta}\left(A-\frac{x}{\delta}\varphi\right)\frac{\sigma_{0}+\sigma_{1}(x/\delta)^{2}}{1+d_{0}(x/\delta)^{2}+d_{1}(x/\delta)^{4}}=\int_{-\infty}^{\infty}dpe^{ipx}\hat{\varphi}(p)F(p\rho_{i})\end{equation}
 and\begin{equation}\label{eq:eq2}
\frac{1}{\hat{\omega}^{2}\hat{\beta}}\frac{d^{2}A}{dx^{2}}=\frac{1}{\delta^{2}}\frac{\delta}{x}\int_{-\infty}^{\infty}dpe^{ipx}\hat{\varphi}(p)F(p\rho_{i}).\end{equation}
Here the electrostatic potential $\Phi$ has been scaled to $\varphi=(\omega L_{s}/k_{y})\Phi,$
$A$ is the parallel component of the vector potential, $x$ is the
distance from the resonant surface, $k_{\parallel}=0,$ and $\hat{\varphi}(p)$
and $\hat{A}(p)$ are the Fourier transform of $\varphi$ and $A.$
In addition we have $\sigma_{0}=1-(1+1.71\eta_{e})/\hat{\omega},$
$\sigma_{1}=d_{1}(1-1/\hat{\omega}),$ $\hat{\omega}=\omega/\omega_{*e},$ $\omega_{*e}=k_yc T_{0e}/(e B_0 a),$ with  $a^{-1}=\partial \ln n_{0e}/\partial x,$
$d_{0}=5.08,$ $d_{1}=2.13,$ \[
\delta^{2}=e^{-i\frac{\pi}{2}}\frac{\omega\nu_{e}}{k_{y}^{2}v_{the}^{2}}L_{s}^{2}\equiv e^{-i\frac{\pi}{2}}\hat{\omega}\delta_{0}^{2},\,\,\,\,\,\,\,\,\hat{\beta}=\frac{\beta_{e}}{2}\frac{L_{s}^{2}}{L_{n}^{2}},\]
 and\begin{equation}
F(p\rho_{i})=\left(\frac{1}{\hat{\omega}}+\tau\right)\left(\Gamma_{0}-1\right)-\frac{1}{\hat{\omega}}\frac{\eta_{i}}{2}p^{2}\rho_{i}^{2}\left(\Gamma_{0}-\Gamma_{1}\right);\,\,\,\,\Gamma_{n}=e^{-\frac{p^{2}\rho_{i}^{2}}{2}}I_{n}(p^{2}\rho_{i}^{2}/2),\end{equation}
with $\rho_{i}^{2}=2m_{i}T_{i}/(e^{2}B_0^{2}).$ Here, $n_{0e}$ and $T_{0e}$ are the equilibrium electron density and temperature, respectively. Equations \eqref{eq:eq1}-\eqref{eq:eq2} are valid when $\omega\sim\omega_{*e}\ll\nu_{e},$ where $\nu_{e}$ is the electron collision frequency, and at the same time $\omega\sim k_{parallel}^2v_{the}^2/\nu_{ei} .$ Thus electrons are {\it not} assumed isothermal and ions are gyro-kinetic.

For our analysis, it is appropriate to
introduce $X=x/\rho_{i}$ and $k=p\rho_{i}.$ Then $F(k)$ has the
following limiting behaviours:

\begin{equation}
F(k)\sim-[\tau+(1+\eta_{i})/\hat{\omega}]k^{2}/2\,\,\,\mbox{for}\,\,\, k\rightarrow0,\end{equation}
\begin{equation}
F(k)\sim F_{\infty}+\frac{f_{1}}{k}\,\,\,\mbox{for}\,\,\, k\rightarrow\infty,\label{eq:asF}\end{equation}
with $F_{\infty}=-(\tau+\hat{\omega}^{-1}),$ $f_{1}=[\tau+\hat{\omega}^{-1}(1-\eta_{i}/2)]/\sqrt{\pi}.$
We solve these equations by asymptotic matching between region $1$,
the ion region, where $x\sim\rho_{i},$ i.e. $X\sim1,$ and region
$2,$ the electron region, where $x\sim\delta,$ i.e. $X\sim\delta/\rho_{i},$
assuming $\rho_{i}\gg\delta$ so that these two regions are widely
separated in $X.$ As noted in the Introduction, the high$-k$ tail accounted for by $f_1/k$ plays a key role in this matching and affects the mode stability \citep{cowley:3230}. 

Following the work of Refs. \citep{pegoraro:478} and \citep{pegoraro:364},
it is convenient to calculate in terms of the current $J,$ where\begin{equation}
J=-\frac{d^{2}A}{dx^{2}},\,\,\,\mbox{or}\,\,\,\,\hat{J}=k^{2}\hat{A},\end{equation}
so that \begin{equation}
\hat{\varphi}=-\frac{i}{\hat{\omega}^{2}\hat{\beta}F(k)}\frac{\delta}{\rho_{i}}\frac{d\hat{J}}{dk}.\end{equation}
 In region $1$ we obtain\begin{equation}
\frac{d}{dk}\left[\frac{G(k)}{F(k)}\frac{d\hat{J}}{dk}\right]+\hat{\omega}^{2}\hat{\beta}\frac{\sigma_{1}}{d_{1}}\frac{\hat{J}}{k^{2}}=0;\,\,\,\, G(k)=F(k)-\frac{\sigma_{1}}{d_{1}}.\label{eq:ion}\end{equation}
Equation \eqref{eq:asF} shows that as $k\rightarrow\infty,$ $F\rightarrow F_{\infty}=-(\tau+\hat{\omega}^{-1})$,
so that\begin{equation}
\hat{J}(k)\sim\hat{a}_{+}k^{1/2+\mu}+\hat{a}_{-}k^{1/2-\mu}\,\,\,\mbox{for\,\,\,}k\rightarrow\infty,\label{eq:largkion}\end{equation}
where $\mu$ is defined by
\begin{equation}
\frac{1}{4}-\mu^{2}=\frac{\hat{\omega}^{2}\hat{\beta}}{1+\tau}\left(\frac{1}{\hat{\omega}}+\tau\right)\left(1-\frac{1}{\hat{\omega}}\right)\end{equation}
 To determine $\hat{a}_{+}/\hat{a}_{-}$ we must solve Eq. \eqref{eq:ion}
and impose the boundary condition arising from the matching to the
ideal region at small $k$ (corresponding to large $x/\rho_{i}$),
given in Ref. \citep{pegoraro:478}:\begin{equation}
\hat{J}(k)\sim1+\frac{\pi}{3\Delta^{\prime}\rho_{i}}\hat{\omega}^{2}\left(\tau+\frac{1+\eta_{i}}{\hat{\omega}}\right)k^{3},\,\,\,\, k\rightarrow0.\label{eq:bcion}\end{equation}
In region $2$ we define $t=\exp(-i\pi/4)(\delta_{0}\hat{\omega}^{1/2}/\rho_{i})k.$
We shall find that it is easier to calculate in $x$ space using the
scaled variable $s=\exp(i\pi/4)(\rho_{i}/\delta_{0}\hat{\omega}^{1/2})X,$
and then transfrom to $t.$ Firstly we obtain\begin{equation}
\left(\sigma_{0}-\sigma_{1}\frac{d^{2}}{dt^{2}}\right)\left(-i\frac{d}{dt}\right)\left(\hat{A}-i\frac{d\hat{\varphi}}{dt}\right)=\left(1-d_{0}\frac{d^{2}}{dt^{2}}+d_{1}\frac{d^{4}}{dt^{4}}\right)F_{\infty}\hat{\varphi},\end{equation}
\begin{equation}
-i\frac{d}{dt}\left(t^{2}\hat{A}\right)=\hat{\omega}^{2}\hat{\beta}F_{\infty}\hat{\varphi},\end{equation}
leading, on integrating once, to\begin{equation}
\left(\sigma_{0}-\sigma_{1}\frac{d^{2}}{dt^{2}}\right)\left[\hat{\omega}^{2}\hat{\beta}\hat{A}+\frac{1}{\tau+\hat{\omega}^{-1}}\frac{d^{2}}{dt^{2}}\left(t^{2}\hat{A}\right)\right]=\left(1-d_{0}\frac{d^{2}}{dt^{2}}+d_{1}\frac{d^{4}}{dt^{4}}\right)\left(t^{2}\hat{A}\right).\label{eq:eltr}\end{equation}
We back-transform this equation to obtain\begin{equation}
\frac{d^{2}A}{ds^{2}}+\bar\sigma\left(\hat s^2\right)\hat{\omega}^{2}\hat{\beta}A=0,\label{eq:electr}\end{equation}
where \begin{equation}\bar\sigma\left(s^2\right)=\frac{\sigma_0+\sigma_1s^2}{1+\bar{d}_0s^2+\bar{d}_1s^4},\end{equation} $\bar{d}_{0}(\hat{\omega})=d_{0}+\sigma_{0}/(\tau+\hat{\omega}^{-1})$ and
 $\bar{d}_{1}(\hat{\omega})=d_{1}+\sigma_{1}/(\tau+\hat{\omega}^{-1})=d_{1}(1+\tau)/(\tau+\hat{\omega}^{-1}).$
An arbitrary constant of integration in Eq. \eqref{eq:eltr} would
lead to an unphysical $\delta-$function at $s=0$ in Eq. \eqref{eq:electr}.
From Eq. \eqref{eq:electr} we write the corresponding equation for $J(s)$
\begin{equation}
\frac{d^{2}}{ds^{2}}\left[\frac{J(s)}{\bar\sigma(s^2)}\right]+\hat{\omega}^{2}\hat{\beta}J=0\label{eq:elecurr}\end{equation}
that must be solved with the condition that $J(s)$ is finite and
even in $s$ at $s=0.$ For large argument, we find\begin{equation}
J(s)\sim b_{+}s^{-3/2+\mu}+b_{-}s^{-3/2-\mu}\,\,\,\mbox{for\,\,\,}s\rightarrow\infty,\label{eq:eq17}\end{equation}
so that for small argument in Fourier space\begin{equation}\label{eq:curraskspace}
\hat{J}(t)\sim\hat{c}_{+}t^{1/2+\mu}+\hat{c}_{-}t^{1/2-\mu}\,\,\,\mbox{for\,\,\,}t\rightarrow0,\end{equation}
where it can be shown that, recalling $J(s)$ is even, \begin{equation}
\frac{\hat{c}_{-}}{\hat{c}_{+}}=\frac{b_{+}}{b_{-}}\frac{\Gamma\left(-\frac{1}{2}+\mu\right)}{\Gamma\left(-\frac{1}{2}-\mu\right)}\tan\left[\frac{\pi}{2}\left(\frac{1}{2}+\mu\right)\right]\label{eq:rule},\end{equation}
which can be matched to the region $1$ solution to yield\begin{equation}
\frac{\hat{c}_{-}}{\hat{c}_{+}}=\frac{\hat{a}_{-}}{\hat{a}_{+}}\left(e^{-i\pi/4}\frac{\hat{\omega}^{1/2}\delta_{0}}{\rho_{i}}\right)^{2\mu}.\label{eq:match}\end{equation}
Thus, to determine $\hat{c}_{-}$ and $\hat{c}_{+},$ we see that
we must solve Eq. \eqref{eq:elecurr} to obtain $b_{-}$ and
$b_{+},$ deduce $\hat{c}_{-}/\hat{c}_{+}$ from Eq. \eqref{eq:rule} which can then be used in the matching condition \eqref{eq:match}
to provide a dispersion relation. This analytic process has fully
accounted for the small parameter $\delta_{0}/\rho_{i}.$ In general,
one needs to solve Eq. \eqref{eq:ion} with boundary condition \eqref{eq:bcion}
to determine $\hat{a}_{-}$ and $\hat{a}_{+}$, then solve Eq. \eqref{eq:elecurr}
with the appropriate boundary condition at $s=0$ to determine $b_{+}$
and $b_{-},$ and hence obtain $\hat{c}_{-}$ and $\hat{c}_{+}$ using
result \eqref{eq:rule}. Normally this would require numerical solutions
of Eqs. \eqref{eq:ion} and \eqref{eq:elecurr}, involving the parameters
$\hat{\beta},$ $\tau,$ $\eta_{e},$ $\eta_{i}$ and $\rho_{i}\Delta^{\prime}.$
However here we shall make some analytic progress in the two limits
$\hat{\beta}\ll1$ and $\hat{\beta}\gg1$, while also obtaining results
valid for finite $\hat{\beta}$ based on the fact that $\hat{\omega}\rightarrow1$
for the tearing mode and $|\hat{\omega}|\ll1$ for the dissipative internal
kink mode at large $\Delta^{\prime}\rho_{i}.$

\section{The Limit $\hat{\beta}\ll1$\label{sec:lowbetaJIM}}

\subsection{A unified dispersion relation at low $\hat{\beta}.$}

Following Ref. \citep{pegoraro:478}, a solution of Eq. \eqref{eq:ion}
for region $1$ satisfying condition \eqref{eq:bcion} at low $k$
can be found by expanding in $\hat{\beta}$ to obtain

\begin{equation}
\begin{split}\hat{J}(k)=\exp\left[\hat{\omega}^{2}\hat{\beta}\frac{\sigma_{1}}{d_{1}}\int_{0}^{k}\frac{du}{u}\frac{F(u)}{G(u)}\right]+\hat{\omega}^{2}\hat{\beta}\frac{\sigma_{1}}{d_{1}}\frac{\pi}{\Delta^{\prime}\rho_{i}}\int_{0}^{k}du\frac{F(u)}{G(u)}\\
-\left(\hat{\omega}^{2}\hat{\beta}\frac{\sigma_{1}}{d_{1}}\right)^{2}\int_{0}^{k}du\frac{F(u)}{G(u)}\int_{0}^{u}dv\frac{F(v)}{v^{2}G(v)}.\end{split}
\label{eq:pegsolions}\end{equation}
At large $k$ this takes the form\begin{equation}
\begin{split} & \hat{J}(k)\sim k^{1/2-\mu}+\frac{\pi}{\Delta^{\prime}\rho_{i}}\left(\frac{1}{4}-\mu^{2}\right)k-\frac{\pi}{\Delta^{\prime}\rho_{i}}\left(\frac{1}{4}-\mu^{2}\right)\frac{\hat{\omega}-1}{1+\tau}\frac{f_{1}}{\hat{\omega}\tau+1}\ln k\\
 & +\frac{\pi}{\Delta^{\prime}\rho_{i}}\left(\frac{1}{4}-\mu^{2}\right)\frac{\hat{\omega}(1+\tau)}{1+\hat{\omega}\tau}\bar{I}(\hat{\omega},\tau,\eta_{i})-\left(\frac{1}{4}-\mu^{2}\right)^{2}\frac{\hat{\omega}(1+\tau)}{1+\hat{\omega}\tau}I(\hat{\omega},\tau,\eta_{i})k\,\,\,\mbox{for\,\,\,}k\rightarrow\infty,\end{split}
\label{eq:ionsollow}\end{equation}
where \begin{equation}
\begin{split} & I(\hat{\omega},\tau,\eta_{i})=\int_{0}^{\infty}dk\frac{F(k)}{k^{2}G(k)}\\
 & \bar{I}(\hat{\omega},\tau,\eta_{i})=\int_{0}^{\infty}dk\left[\frac{F(k)}{G(k)}-\frac{1+\hat{\omega}\tau}{\hat{\omega}(1+\tau)}+\frac{(\hat{\omega}-1)(1+\hat{\omega}\tau-\eta_{i}/2)}{\sqrt{\pi}\hat{\omega}^{2}(1+\tau)^{2}(1+k)}\right],\end{split}
\end{equation}
where we have recalled the asymptotic form of $F(k)$ at large $k$
given in Eq. \eqref{eq:asF}. We note that the term involving $f_1$ gives rise
to the $\ln k$ term in Eq. (\ref{eq:ionsollow}).

In region $2$ we solve Eq. \eqref{eq:elecurr} to zeroth order in
$\hat{\beta}$ to obtain\begin{equation}
J^{(0)}(s)=J_0\bar\sigma(s^2),\label{eq:zerothel}\end{equation} where $J_0$ is a constant. In
next order we find \begin{equation}J^{(1)}(s)=-\hat\omega^2\hat\beta J^{(0)}\int_0^sds^{\prime}\left(s-s^{\prime}\right)\bar\sigma\left({s^{\prime}}^2\right).\end{equation}
At large $s,$ evaluating the integral $I_e=\int_0^{\infty}\bar\sigma\left(s^2\right)$ by contour integration, we obtain \begin{equation}\label{eq:lowbetacurrexp}
J(s) \sim J_0\frac{\sigma_{1}}{\bar{d}_{1}s^{2}}\left\{ 1-\frac{\pi}{2\bar{d}_{1}}\frac{\hat\omega^2\hat\beta}{(s_{+}+s_{-})}\left(\frac{\sigma_{0}}{s_{+}s_{-}}-\sigma_{1}\right)s+\hat\omega^2\hat\beta\frac{\sigma_{1}}{\bar{d}_{1}}\ln s^2\right\},\,\,\,\mbox{for}\,s\rightarrow\infty,  
\end{equation}
where \begin{equation}
s_{\pm}^{2}=\frac{\bar{d}_{0}}{2\bar{d}_{1}}\pm\sqrt{\frac{\bar{d}_{0}^{2}}{4\bar{d}_{1}^{2}}-\frac{1}{\bar{d}_{1}}}.\end{equation}
At low $\hat\beta$ we must take the limit $\mu\rightarrow 1/2$ of $J(s)$ in Eq.\eqref{eq:eq17}:
\begin{equation}\label{eq:currasklow}
J(s)\sim\frac{1}{s^{2}}\left\{ b_{-}\left(1+\hat{\omega}^{2}\hat{\beta}\frac{\sigma_{1}}{\bar{d}_{1}}\ln s^2+\ldots\right)+b_{+}\left(s+\ldots\right)\right\},\,\,\,\mbox{for}\,s\rightarrow\infty, 
\end{equation} so that we can identify $b_+$ and $b_{-}$ by comparison with Eq. \eqref{eq:lowbetacurrexp}:
\begin{equation}
\frac{b_+}{b_-}=-\hat\omega^2\hat\beta\frac{\pi}{2\bar d_1}\frac{1}{s_++s_-}\left(\frac{\sigma_0}{s_+s_-}-\sigma_1\right).
\end{equation}
Then Eq. \eqref{eq:match} in the limit $\mu\rightarrow 1/2$ implies
\begin{equation}\label{eq:cpcmlow}\frac{\hat c_-}{\hat c_+}=\frac{2}{\pi}\frac{\bar d_1}{\sigma_1}\frac{1}{\hat\omega^2\hat\beta} \frac{b_+}{b_-},\end{equation} so that
\begin{equation}\label{eq:currlowmatch}
\hat J(t)\approx \hat c_- t^{1/2-\mu}+\hat c_+ t,\,\,\,\mbox{for}\,t\rightarrow 0.
\end{equation}

Matching the powers in the solutions given by Eqs. \eqref{eq:ionsollow}
and \eqref{eq:currlowmatch}, with the use of Eq. \eqref{eq:cpcmlow}, yields the dispersion relation\begin{equation}
\begin{split} & e^{-i\frac{\pi}{4}}\frac{\delta_{0}}{\rho_{i}}\frac{\hat{\omega}^{1/2}\sigma_{1}}{\sqrt{\bar{d}_{1}}}\frac{\sqrt{\bar{d}_{0}+2\sqrt{\bar{d}_{1}}}}{\sigma_{0}\sqrt{\bar{d}_{1}}+\sigma_{1}}=\\
 & -\frac{\frac{1+\hat{\omega}\tau}{1+\tau}(\hat{\omega}-1)\left[\frac{\pi\hat{\beta}}{\Delta^{\prime}\rho_{i}}-\hat{\omega}(\hat{\omega}-1)\hat{\beta}^{2}I\right]}{1-\hat{\omega}(\hat{\omega}-1)\frac{\pi\hat{\beta}}{\Delta^{\prime}\rho_{i}}\left[\frac{(\hat{\omega}-1)(\hat{\omega}\tau+1-\eta_{i}/2)}{\sqrt{\pi}(1+\tau)^{2}\hat{\omega}^{2}}\ln\left(e^{i\frac{\pi}{4}}\frac{\rho_{i}}{\delta_{0}\hat{\omega}^{1/2}}\right)-\bar{I}\right]}.\end{split}
\label{eq:displow}\end{equation}
To analyse Eq. \eqref{eq:displow} it is useful to cast it in the
form \begin{equation}
e^{-i\frac{\pi}{4}}\frac{\delta_{0}}{\rho_{i}}A(\hat{\omega})B(\hat{\omega})+\hat{\omega}\sqrt{1+\hat{\omega}\tau}C(\hat{\omega})D(\hat{\omega})=0,\label{eq:dispJim}\end{equation}
where we have defined\begin{equation}
\begin{split} & A(\hat{\omega})=\sqrt{1+\tau}\sqrt{\hat{\omega}-(1+1.71\eta_{e})+d_{0}(1+\hat{\omega}\tau)+2\sqrt{d_{1}(1+\tau)(1+\hat{\omega}\tau)\hat{\omega}}}\\
 & B(\hat{\omega})=\frac{\Delta^{\prime}\rho_{i}}{\pi\hat{\beta}}\hat{\omega}-(\hat{\omega}-1)^{2}\frac{(\hat{\omega}\tau+1-\eta_{i}/2)}{\sqrt{\pi}(1+\tau)^{2}}\ln\left(e^{i\frac{\pi}{4}}\frac{\rho_{i}}{\delta_{0}\hat{\omega}^{1/2}}\right)+\hat{\omega}^{2}(\hat{\omega}-1)\bar{I}(\hat{\omega},\tau,\eta_{i})\\
 & C(\hat{\omega})=1-\frac{\Delta^{\prime}\rho_{i}\hat{\beta}}{\pi}\hat{\omega}(\hat{\omega}-1)I(\hat{\omega},\tau,\eta_{i})\\
 & D(\hat{\omega})=[\hat{\omega}-(1+1.71\eta_{e})]\sqrt{\hat{\omega}(1+\tau)}+(\hat{\omega}-1)\sqrt{d_{1}(1+\hat{\omega}\tau)}.\end{split}
\end{equation}
The important $\ln(\rho_i/\delta_0)$ term in $B(\hat\omega)$ arises from the $f_1/k$ correction in Eq. (\ref{eq:asF}). Since $\delta_{0}/\rho_{i}\ll1$, there appear to be four uncoupled
solution branches for the lowest order eigenvalue, $\hat{\omega}_{0},$
given by\begin{equation}
\hat{\omega}_{0}=-1/\tau,\,\,\, C(\hat{\omega}_{0})=0,\,\,\,\mbox{and}\,\,\, D(\hat{\omega}_{0})=0.\label{eq:solfund}\end{equation}
The first of these is an ion drift mode. The fourth branch, $D(\hat{\omega}_{0})=0,$
is the drift-tearing mode whose stability is discussed shortly. The
other two branches arise from solutions of $C(\hat{\omega}_{0})=0,$
corresponding to a pair of kinetic Alfv\'en waves (KAW's) if $\Delta^{\prime}>0.$
To see this we consider the limit $|\hat{\omega}|\gg1,$ when $I(\hat{\omega})\approx0.798+(1.194+0.399\eta_{i})/\hat{\omega},$
where we have taken the limit $\tau=1$ for simplicity. Then we find
that\begin{equation}
\hat{\omega}=\pm\sqrt{\frac{\pi}{0.8\Delta^{\prime}\rho_{i}\hat{\beta}}},\end{equation}
where we require $\hat\beta^{-1}\gg \Delta^{\prime}\rho_i\gg 1,$ so that the conditions
$\hat\omega^2\ll 1$ and $\hat\omega\gg 1$ are met.
In Alfv\'enic units, the frequency of the mode is \begin{equation}
\omega=\pm\omega_{KAW}\equiv\pm\sqrt{\frac{\pi}{0.8\Delta^{\prime}\rho_{i}}}k_{y}\rho_{i}\frac{v_{A}}{L_{s}},\label{eq:KAW}\end{equation}
where we have defined the Alfv\'en speed $v_{A}=B/\sqrt{m_{i}n_{i}}.$
The two solutions in Eq. \eqref{eq:KAW} correspond to one KAW propagating
in the electron diamagnetic direction and another in the ion direction. 

\begin{figure}[h]
\includegraphics[height=0.26\textheight]{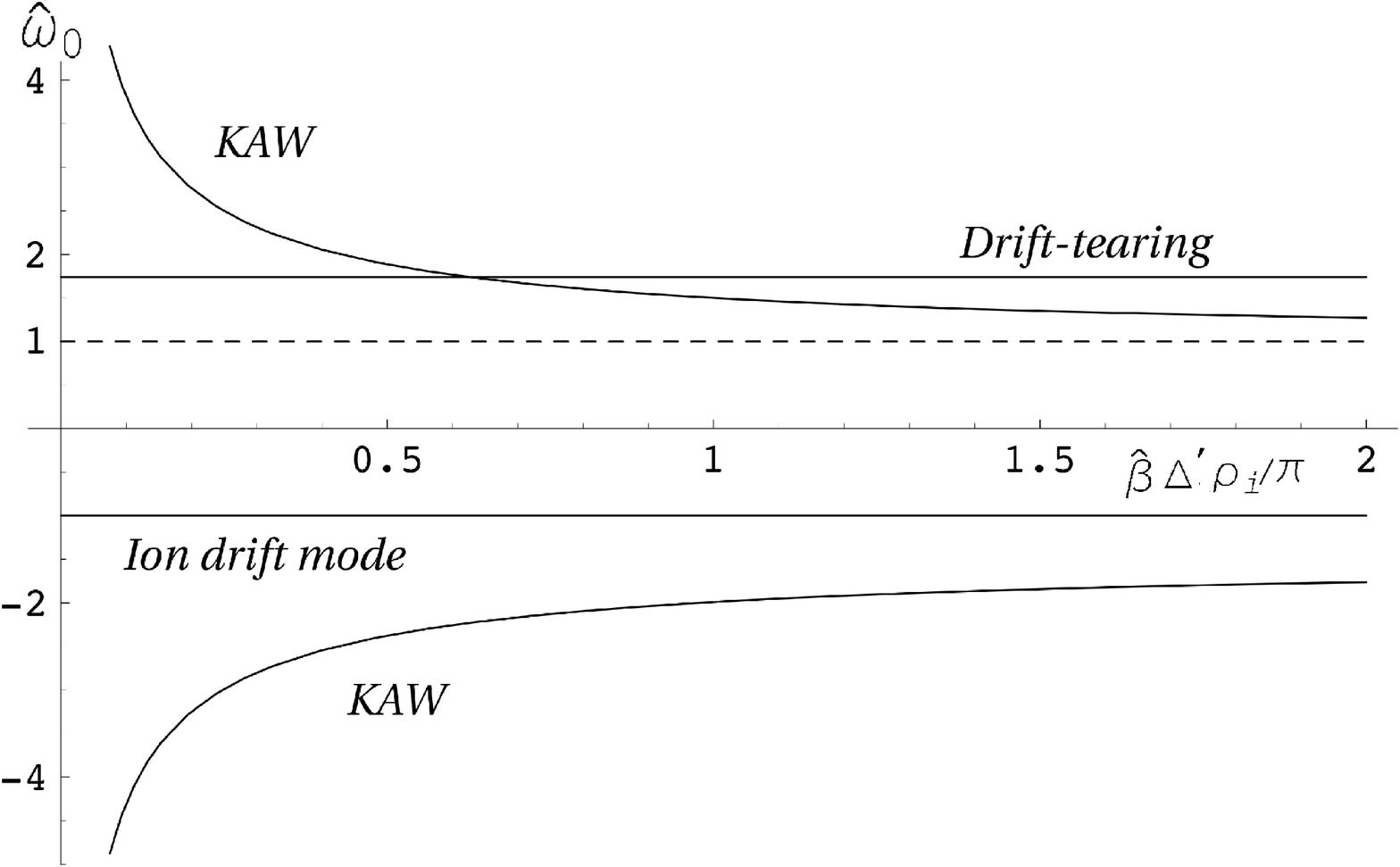} 

\caption{The solutions of the dispersion equation \eqref{eq:dispJim} for the
drift-tearing mode and two KAWs as a function of $\Delta^{\prime}\rho_{i}\hat{\beta}/\pi.$
There is also an ion drift mode. The drift-tearing mode and
the KAW propagating in the electron direction are strongly coupled
in the vicinity of $\Delta^{\prime}\rho_{i}\hat{\beta}=(\Delta^{\prime}\rho_{i}\hat{\beta})_{c}$
given in Eq. \eqref{eq:cross-crit}.}
\label{fig:fig1}
\end{figure}
 Including the $\delta_{0}/\rho_{i}$ correction to Eq. \eqref{eq:dispJim},
we find that these two KAWs are stable. The lowest order drift-tearing
and KAW mode frequencies as functions of $\Delta^{\prime}\rho_{i}\hat{\beta}$
are shown schematically in Fig. \ref{fig:fig1}. At large values of
$\Delta^{\prime},$ a further solution of $C(\hat{\omega})=0$ is
associated with the dissipative internal kink mode discussed in Section
\ref{sec:The-Internal-Kink}, while negative values of $\Delta^{\prime}$
give rise to the FLR modified ideal internal kink mode analysed in
Ref. \citep{pegoraro:478}.

However we note that at small $\hat{\beta}$ when $\delta_{0}/\rho_{i}\sim\hat{\beta}^{2},$
but $\Delta^{\prime}\rho_{i}\hat{\beta}\sim\mathcal{O}(1),$ which
introduces an interaction between the drift-tearing mode and the KAWs,
all four branches become coupled and Eq. \eqref{eq:dispJim} no longer
has a natural expansion parameter.

\subsection{The drift-tearing mode}\label{subsec:lowbetatearing}

Returning to the drift-tearing mode and assuming $\Delta^{\prime}\rho_{i}\hat{\beta}\ll1$
so that we have an uncoupled mode, the lowest order solution, $D(\hat{\omega})=0,$
is given by \begin{equation}
\left(1-\frac{1+1.71\eta_{e}}{\hat{\omega}}\right)\left(\frac{1+\tau}{\tau+\hat{\omega}^{-1}}\right)^{1/2}+1.45\left(1-\frac{1}{\hat{\omega}}\right)=0.\label{eq:dtreal}\end{equation}
\begin{figure}[h]
\includegraphics[height=0.26\textheight]{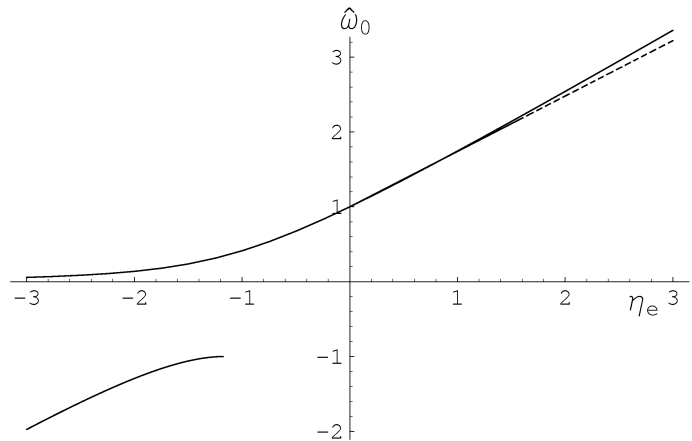} 

\caption{The drift-tearing mode frequency {[}the solution of Eq. \eqref{eq:dtreal}]
as a function of $\eta_{e}$ for $\tau=1.$ For negative $\eta_{e}<-1.17$
a second solution exists. Also shown as a dotted line is the fit \eqref{eq:realdt}.}
\label{fig:fig2}
\end{figure}
 The solution of Eq. \eqref{eq:dtreal} is shown in Fig. \ref{fig:fig2};
for $\eta_{e}>0$ (and $\tau=1$), a good fit to the solution is given
by \begin{equation}
\hat{\omega}=\hat{\omega}_{0}\approx1+0.74\eta_{e}.\label{eq:realdt}\end{equation}
Figure \ref{fig:fig2} also shows a second solution for negative $\eta_{e},$ i.e. reversed density gradients or hollow temperature profiles;
in fact it exists for $\eta_{e}<-(1+\tau^{-1})/1.71,$ i.e. $\eta_{e}<-1.17$
for $\tau=1,$ and for larger negative $\eta_{e}$ it can also be
fitted by the form \eqref{eq:realdt}. Thus, in the limit of a flat
density profile, one does obtain the same frequency, irrespective
of the sign of $\eta_{e}.$ The growth rate follows from next order
in $\delta_{0}/\rho_{i}:$ \begin{equation}
\hat\gamma\equiv\frac{\gamma}{\omega_{*e}}\sim\frac{\delta_{0}\hat{\omega}_{0}^{1/2}}{\pi\hat{\beta}}\left(\Delta^{\prime}-\Delta_{crit}^{\prime}\right),\label{eq:dtgrowth}\end{equation}
where $\Delta_{crit}^{\prime}$ is the critical $\Delta^{\prime}$ for 
instability\begin{equation}
\Delta_{crit}^{\prime}=\frac{\sqrt{\pi}\hat{\beta}}{\rho_{i}}\frac{1+\hat{\omega}_{0}\tau}{\hat{\omega}_{0}^{2}(1+\tau)^{2}}\left(\hat{\omega}_{0}\tau+1-\frac{\eta_{i}}{2}\right)(\hat{\omega}_{0}-1)^{2}\ln\left(e^{-\frac{\pi}{4}}\frac{\rho_{i}}{\delta_{0}\hat{\omega}_{0}^{1/2}}\right)-\frac{\pi\hat{\beta}}{\rho_{i}}(\hat{\omega}_{0}-1)\bar{I},\label{eq:critdelprime}\end{equation}
provided $\eta_{i}/2<\hat{\omega}_{0}\tau+1,$ where we note that
expression \eqref{eq:realdt} implies the factor $(\hat{\omega}_{0}-1)^{2}\propto\eta_{e}^{2}.$
This result closely resembles that obtained in Ref. \citep{cowley:3230}.
The integral $\bar{I}[\hat \omega_0(\eta_{e}),\eta_{i},\tau],$ which is negative,
scales approximately as $\bar{I}\approx\eta_{e}^{1/2},$ so the stabilising
effect of this term resembles the diamagnetic stabilisation found
in Ref. \citep{drake:2509}:\begin{equation}
\Delta_{crit}^{\prime dia}=\frac{\pi\hat{\beta}}{\rho_{i}}(\hat{\omega}_{0}-1)\left|\bar{I}\right|\label{eq:diamcritdel}\end{equation}
as can be seen by taking the $\tau\rightarrow\infty$ limit of Eq.
\eqref{eq:diamcritdel}, noting that $\bar{I}\sim\tau^{-1/2}$ in
this situation. %
\begin{figure}[h]
\includegraphics[height=0.26\textheight]{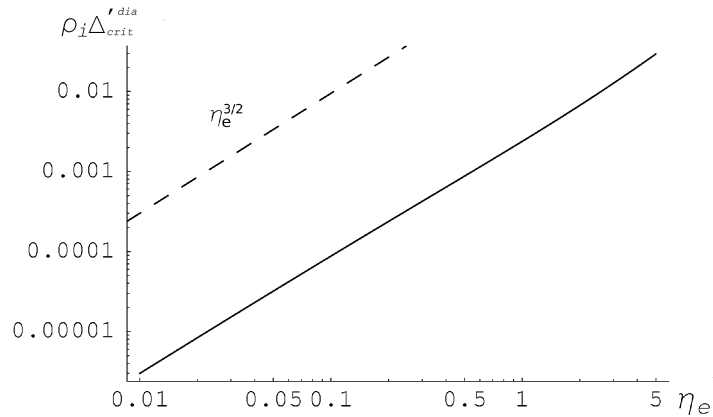} 

\caption{The scaling of the quantity $\Delta_{crit}^{\prime dia}$ in Eq. \eqref{eq:diamcritdel}}
\label{fig:fig4}
\end{figure}
 The quantity $\Delta_{crit}^{\prime dia}$ is shown as a function
of $\eta_{e}$ in Fig. \ref{fig:fig4}. The two different effects
are shown in Fig. \ref{fig:fig3}, %
\begin{figure}[h]
\includegraphics[height=0.26\textheight]{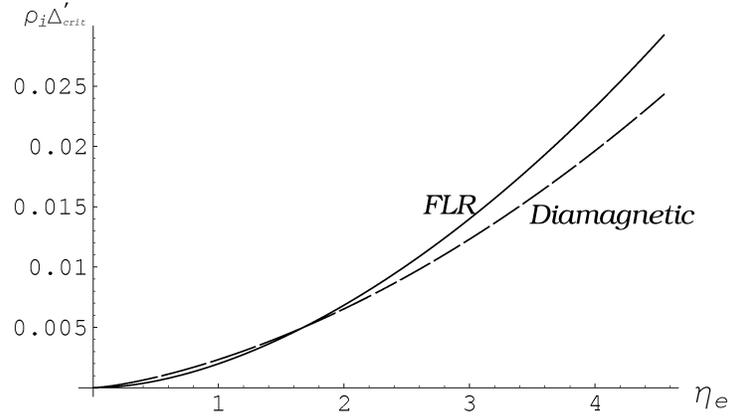} 

\caption{The diamagnetic and ion-kinetic contributions to the critical $\Delta^{\prime}$
[given in Eq. \eqref{eq:diamcritdel} and \eqref{eq:FLRcritdel}, respectively] as a function of the
electron temperature gradients $\eta_{e}.$ Here $\hat{\beta}=10^{-3},$
$\delta_{0}/\rho_{i}=10^{-3},$ $\tau=1,$ $\eta_{i}=0.$}
\label{fig:fig3}
\end{figure}
where \begin{equation}\label{eq:FLRcritdel}\Delta^{\prime\,FLR}_{crit}=\frac{\sqrt{\pi}\hat{\beta}}{\rho_{i}}\frac{1+\hat{\omega}_{0}\tau}{\hat{\omega}_{0}^{2}(1+\tau)^{2}}\left(\hat{\omega}_{0}\tau+1-\frac{\eta_{i}}{2}\right)(\hat{\omega}_{0}-1)^{2}\ln\left(e^{-\frac{\pi}{4}}\frac{\rho_{i}}{\delta_{0}\hat{\omega}_{0}^{1/2}}\right).\end{equation}
 It should also be noted that, in principle, a large enough \textsl{ion}
temperature gradient ($\eta_{i}/2>\hat{\omega}_{0}\tau+1$) can give
a \textsl{negative} critical delta prime, however this requires values
of $\eta_{i}$ that are probably only physically relevant in the flat density limit.

When $\eta_{e}<0$ we have seen there are solutions of Eq. \eqref{eq:dtreal}
with $\hat{\omega}<1,$ in which case the integrand over $k$ in the
integral $\bar{I}(\hat{\omega})$ has a pole at $k=k_{0},$ where
$G(k_{0})=0.$ This zero in $G(k)$ corresponds to a resonance with
an electrostatic drift wave of wave-number $k.$ The integral $\bar{I}(\hat{\omega})$
then has an imaginary contribution: by considering $\hat{\omega}$
to have a small imaginary part, we find the integral along real $k$
must pass above the pole at $k=k_{0}$ to ensure causality. This contribution
provides a stabilising effect that increases the value of $\Delta_{crit}^{\prime}$
in Eq. \eqref{eq:critdelprime} via damping on the electrostatic electron
drift wave.

\subsection{The kinetic Alfv\'en wave.}\label{subsec:KAW}

According to Eq. \eqref{eq:dtgrowth}, as the tearing parameter $\Delta^{\prime}\rho_{i}/(\pi\hat{\beta})$
is increased beyond its marginal stability value \eqref{eq:critdelprime},
the drift-tearing mode becomes increasingly unstable. However, when
$\Delta^{\prime}\rho_{i}\hat{\beta}\sim\mathcal{O}(1)$ the approximation
$C(\hat{\omega}_{0})\approx1$ fails; the dispersion relation can
then be written in the form\begin{equation}
e^{-i\frac{\pi}{4}}\frac{\delta_{0}}{\rho_{i}\hat{\beta}^{2}}A(\hat{\omega})\frac{\Delta^{\prime}\rho_{i}\hat{\beta}}{\pi}+\hat{\omega}\sqrt{1+\hat{\omega}\tau}\left[1-\frac{\Delta^{\prime}\rho_{i}\hat{\beta}}{\pi}\hat{\omega}(\hat{\omega}-1)I(\hat{\omega})\right]D(\hat{\omega})=0.\label{eq:KAWdisp}\end{equation}
At the drift-tearing frequency, $\hat{\omega}_{0}\approx1+0.74\eta_{e},$
the ion integral $\bar{I}$ is real and positive so the KAW propagating
in the electron direction becomes strongly coupled to the drift-tearing
branch when $\Delta^{\prime}\rho_{i}\hat{\beta}\sim\mathcal{O}(1).$
At yet higher values of this parameter the modes again decouple. We
shall show below that the KAW propagating in the electron drift direction
connects with the tearing mode having the frequency $\hat{\omega}_{0}.$
Because $C(\hat{\omega}_{0})$ has now changed sign, the branch with $\omega=\hat\omega_0$ has become stable. On the other hand the drift-tearing mode couples to
the KAW given by the solution of $C(\hat{\omega}_{0})=0.$ Since one
can show that, near $\hat{\omega}=1,$ \begin{equation}
I(\hat{\omega}_{0})\approx\frac{\pi}{2}\sqrt{\frac{(1+\tau+\eta_{i})/2}{\hat{\omega}-1}},\end{equation}
this branch has an unstable solution with \begin{equation}
\hat{\omega}=1+\frac{2}{1+\tau+\eta_{i}}\left(\frac{2}{\Delta^{\prime}\rho_{i}\hat{\beta}}\right)^{2}-\frac{8\sqrt{2}\sqrt{d_{0}+2\sqrt{d_{1}}-1.71\eta_e/(1+\tau)}}{1.71\eta_{e}(1+\tau+\eta_{i})}\frac{1}{\hat{\beta}^{3}\Delta^{\prime}\rho_{i}}\frac{\delta_{0}}{\rho_{i}},\label{eq:realKAW}\end{equation}
and a growth rate\begin{equation}
\begin{split} & \hat{\gamma}=\frac{16}{\sqrt{2}\pi}\frac{\delta_{0}}{\rho_{i}}\frac{\sqrt{d_{0}+2\sqrt{d_{1}}-1.71\eta_e/(1+\tau)}}{\hat{\beta}^{3}\Delta^{\prime}\rho_{i}1.71\eta_{e}(1+\tau+\eta_{i})}\times\\
 & \left[1-\frac{1}{\sqrt{2}\pi}\frac{\delta_{0}}{\rho_{i}}\frac{\Delta^{\prime}\rho_{i}}{\hat{\beta}}\frac{\sqrt{d_{0}+2\sqrt{d_{1}}-1.71\eta_e/(1+\tau)}}{1.71\eta_{e}}\right].\end{split}
\label{eq:growthKAW}\end{equation}
Thus, as $\hat{\beta}$ increases, $\hat{\omega}\rightarrow1$ and
the growth rate decreases to zero. In fact, at very high values of
$\Delta^{\prime}\rho_{i}$ the mode becomes stable. The condition
for this is \begin{equation}
\Delta^{\prime}\rho_{i}>2.42\pi\frac{\rho_{i}}{\delta_{0}}\frac{\eta_{e}\hat{\beta}}{\sqrt{d_{0}+2\sqrt{d_{1}}-1.71\eta_e/(1+\tau)}}.\label{eq:stabKAW}\end{equation}
This stabilisation mechanism does not exist when $\hat\beta<{(\delta_0/\rho_i)}^{1/2}$ since the perturbative solution 
of Eq.\eqref{eq:KAWdisp} fails for $\Delta^{\prime }\rho_i \hat\beta\sim\mathcal O (1).$ 
Numerical solution shows that the unstable drift-tearing mode persists as $\Delta^{\prime}\rho_i$ increases
beyond the critical value given in criterion \eqref{eq:critdelprime}, eventually merging with the dissipative 
kink instability discussed later in Section \ref{sec:The-Internal-Kink} when $\Delta^{\prime}\rho_i\gg1.$

We also note that Eq. \eqref{eq:growthKAW} implies stability when\begin{equation}
d_{0}+2\sqrt{d_{1}}-1.71\eta_{e}/(1+\tau)=0,\end{equation}\label{eq:stab_KAW_new}
corresponding to a critical $\eta_{e}.$ Hence, as $\hat{\omega}\rightarrow1,$
the drift tearing mode is also stable if \begin{equation}
\eta_{e}>4.68(1+\tau).\end{equation}

\subsection{The coupled modes}

\begin{figure}[h]
\includegraphics[height=0.26\textheight]{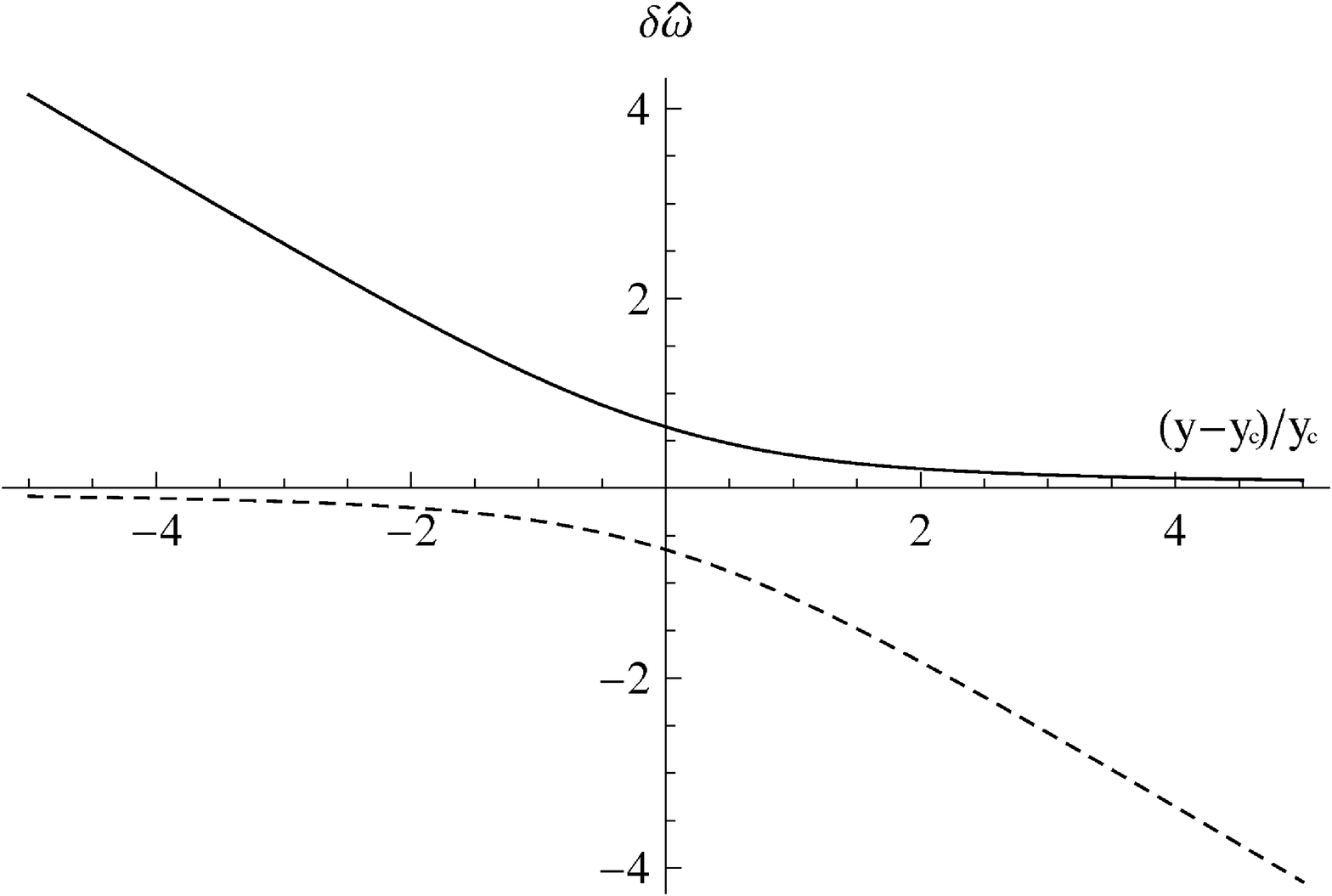} 

\caption{The variation of the real frequencies of the two modes with $\Delta^{\prime}\rho_{i}\hat{\beta},$
for fixed $\hat{\beta}=0.2$ and $\delta_{0}/\rho_{i}=0.1,$ in the
vicinity of the cross-over point, $y\equiv\Delta^{\prime}\rho_{i}\hat{\beta}=(\Delta^{\prime}\rho_{i}\hat{\beta})_{c}\equiv y_{c}$
and $\hat{\omega}=\hat{\omega}_{0},$ showing how the KAW converts
to the drift-tearing mode and vice versa. These results are for a
typical case: $\eta_{e}=\eta_{i}=\tau=1.$}
\label{fig:fig5}
\end{figure}

To analyse how the two branches interact we examine the solution of
the dispersion relation \eqref{eq:KAWdisp} in the vicinity of the
cross-over point. At this point, $C(\hat{\omega})=D(\hat{\omega})=0,$
so that $\hat{\omega}$ is given by the solution of Eq. \eqref{eq:dtreal},
i.e. $\hat{\omega}=\hat{\omega}_{0},$ and the critical value of the
parameter $\Delta^{\prime}\rho_{i}\hat{\beta}$ is given by\begin{equation}
(\Delta^{\prime}\rho_{i}\hat{\beta})_{c}=\frac{\pi}{\hat{\omega}_{0}(\hat{\omega}_{0}-1)I(\hat{\omega}_{0},\eta_{i},\tau)},\label{eq:cross-crit}\end{equation}
where the integral $I$ is real and positive for modes propagating
in the electron direction. Here the two modes interact strongly. After introducting $\delta\hat\omega=\hat\omega-\hat\omega_0$ and
$\delta(\Delta^{\prime}\rho_i\hat\beta)=\Delta^{\prime}\rho_i\hat\beta-(\Delta^{\prime}\rho_i\hat\beta)_c,$
a perturbation theory yields\begin{equation}
(\delta\hat{\omega})^{2}\left(\frac{\partial C}{\partial\hat{\omega}}\frac{\partial D}{\partial\hat{\omega}}\right)_{\hat{\omega}_{0}}-\delta\hat{\omega}\left(\frac{\partial D}{\partial\hat{\omega}}\right)_{\hat{\omega}_{0}}\frac{\delta(\Delta^{\prime}\rho_{i}\hat{\beta})}{(\Delta^{\prime}\rho_{i}\hat{\beta})_{c}}+e^{-i\frac{\pi}{4}}\frac{\delta_{0}}{\rho_{i}}\frac{A(\hat{\omega}_{0})B(\hat{\omega}_{0})}{\hat{\omega}_{0}\sqrt{1+\hat{\omega}_{0}\tau}}=0.\label{eq:coupled}\end{equation}
One can trace how the branches reconnect about this point using the
solution of Eq. \eqref{eq:coupled}:\begin{equation}
\begin{split} & \delta\hat{\omega}_{\pm}=\frac{1}{2(\partial C/\partial\hat{\omega})_{\hat{\omega}_{0}}}\frac{\delta(\Delta^{\prime}\rho_{i}\hat{\beta})}{(\Delta^{\prime}\rho_{i}\hat{\beta})_{c}}\pm\frac{1}{2(\partial C/\partial\hat{\omega})_{\hat{\omega}_{0}}}\times\\
 & \sqrt{\left[\frac{\delta(\Delta^{\prime}\rho_{i}\hat{\beta})}{(\Delta^{\prime}\rho_{i}\hat{\beta})_{c}}\right]^{2}-4e^{-i\frac{\pi}{4}}\frac{\delta_{0}}{\rho_{i}}\frac{(\partial C/\partial\hat{\omega})_{\hat{\omega}_{0}}}{(\partial D/\partial\hat{\omega})_{\hat{\omega}_{0}}}\frac{A(\hat{\omega}_{0})B(\hat{\omega}_{0})}{\hat{\omega}_{0}\sqrt{1+\hat{\omega}_{0}\tau}}}.\end{split}
\label{eq:solcoupled}\end{equation}
When \begin{equation}
\left[\frac{\delta(\Delta^{\prime}\rho_{i}\hat{\beta})}{(\Delta^{\prime}\rho_{i}\hat{\beta})_{c}}\right]^{2}\gg\left|4e^{-i\frac{\pi}{4}}\frac{\delta_{0}}{\rho_{i}}\frac{(\partial C/\partial\hat{\omega})_{\hat{\omega}_{0}}}{(\partial D/\partial\hat{\omega})_{\hat{\omega}_{0}}}\frac{A(\hat{\omega}_{0})B(\hat{\omega}_{0})}{\hat{\omega}_{0}\sqrt{1+\hat{\omega}_{0}\tau}}\right|,\end{equation}
solutions \eqref{eq:solcoupled} simplify to\begin{equation}
\begin{split} & \delta\hat{\omega}_{\pm}=\frac{1}{2(\partial C/\partial\hat{\omega})_{\hat{\omega}_{0}}}\frac{\delta(\Delta^{\prime}\rho_{i}\hat{\beta})}{(\Delta^{\prime}\rho_{i}\hat{\beta})_{c}}\pm\left|\frac{1}{2(\partial C/\partial\hat{\omega})_{\hat{\omega}_{0}}}\frac{\delta(\Delta^{\prime}\rho_{i}\hat{\beta})}{(\Delta^{\prime}\rho_{i}\hat{\beta})_{c}}\right|\\
 & \mp e^{-i\frac{\pi}{4}}\frac{\delta_{0}}{\rho_{i}}\frac{A(\hat{\omega}_{0})B(\hat{\omega}_{0})}{\hat{\omega}_{0}\sqrt{1+\hat{\omega}_{0}\tau}}\left[\frac{1}{2(\partial D/\partial\hat{\omega})_{\hat{\omega}_{0}}}\frac{\delta(\Delta^{\prime}\rho_{i}\hat{\beta})}{(\Delta^{\prime}\rho_{i}\hat{\beta})_{c}}\right]^{-1}\mbox{sign}\left[(\partial C/\partial\hat{\omega})_{\hat{\omega}_{0}}\right].\end{split}
\label{eq:cop}\end{equation}
Thus we see that for $\Delta^{\prime}\rho_{i}\hat{\beta}<(\Delta^{\prime}\rho_{i}\hat{\beta})_{c},$
$\delta\hat{\omega}_{+}$ connects to the stable KAW branch, while
for $\Delta^{\prime}\rho_{i}\hat{\beta}>(\Delta^{\prime}\rho_{i}\hat{\beta})_{c}$
it connects to the stable drift-tearing mode, whereas $\delta\hat{\omega}_{-}$
connects to the unstable drift-tearing mode for $\Delta^{\prime}\rho_{i}\hat{\beta}<(\Delta^{\prime}\rho_{i}\hat{\beta})_{c}$
and to the unstable KAW for $\Delta^{\prime}\rho_{i}\hat{\beta}>(\Delta^{\prime}\rho_{i}\hat{\beta})_{c}.$
This is shown in Fig. \ref{fig:fig5}, %
\begin{figure}[h]
\includegraphics[height=0.26\textheight]{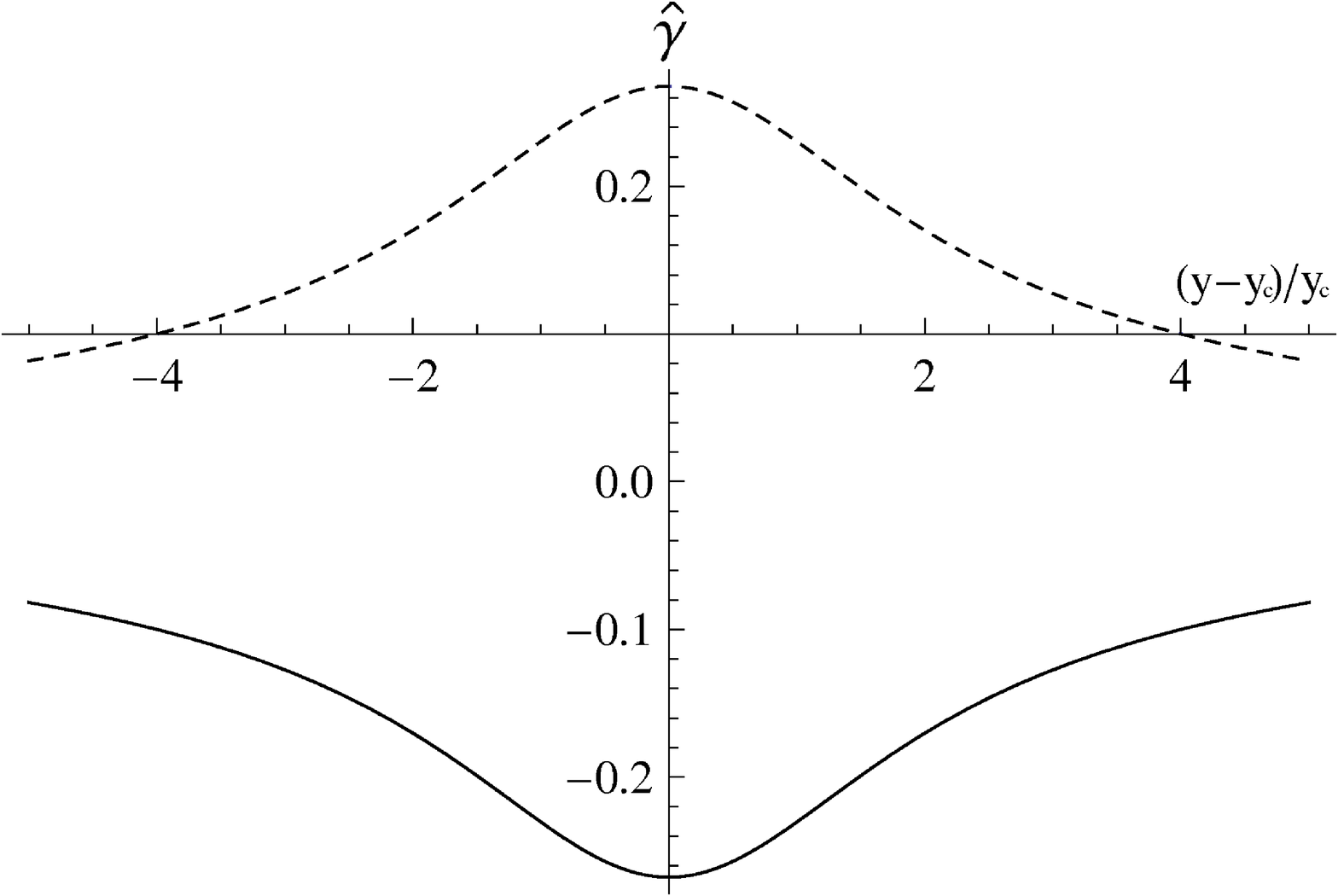} 

\caption{The equivalent variation of the growth/damping rates of the two modes
in Fig. \ref{fig:fig5}, showing how they peak at the cross-over point.
These results are for $\eta_{e}=\eta_{i}=\tau=1.$}
\label{fig:fig6}
\end{figure}
 where we plot $\delta\hat{\omega}_{\pm}$ in the vicinity of the
cross-over, by varying the parameter $\Delta^{\prime}\rho_{i}\hat{\beta}$
at constant $\hat{\beta}=0.2,$ for $\delta_{0}/\rho_{i}=0.1,$ and
$\eta_{e}=\eta_{i}=\tau=1.$ {[}the quantities in Eq. \eqref{eq:cop}
are very insensitive to these latter three parameters]. Very close
to $\Delta^{\prime}\rho_{i}\hat{\beta}=(\Delta^{\prime}\rho_{i}\hat{\beta})_{c}$
the unstable branch has a larger growth rate than is given by Eq.
\eqref{eq:growthKAW}, namely \begin{equation}
\hat{\gamma}\sim\sqrt{\frac{\delta_{0}}{\rho_{i}}}.\end{equation}
In Fig. \ref{fig:fig6} we show the dependence of the growth (or damping)
rates of the two modes for the same parameters as in Fig. \ref{fig:fig5}.

\section{The limit $\hat{\beta}\gg1$\label{sec:highbeta}}

The analysis in the previous section was limited to the case $\hat{\beta}\ll1;$
in this Section we explore the opposite case: $\hat{\beta}\gg1.$
We anticipate that the ``intermediate eigenvalue'' $\mu$ will be finite, so
that at high $\hat{\beta}$ we will have $\hat{\omega}-1\sim\mathcal{O}(\hat{\beta}^{-1}),$
as in Ref. \citep{drake:2509}. However, we have also seen in the
low $\hat{\beta}$ theory of Section \ref{sec:lowbetaJIM} that $\hat{\omega}\rightarrow1$
when $\Delta^{\prime}\rho_{i}\hat{\beta}\gg1.$ In Section \ref{sec:finitebeta}
we will consider the fact that we can also have $\hat{\omega}-1\sim\varepsilon\ll1$
at finite $\hat{\beta}$ to provide a bridge between the high and
low $\hat{\beta}$ limits.

If $\hat{\omega}-1\sim\mathcal{O}(\hat{\beta}^{-1}),$ then $\sigma/d_{1}F(k)\ll1$
in Eq. \eqref{eq:ion} for region $1,$ unless $k\ll1,$ in which
case $F(k)\propto k^{2}.$Thus for finite $k$ {[}not just for $k\gg1$,
as in Eq. \eqref{eq:largkion}] the solution of Eq. \eqref{eq:ion}
is \begin{equation}
\hat{J}(k)=\hat{a}_{+}k^{1/2+\mu}+\hat{a}_{-}k^{1/2-\mu}.\end{equation}
For $k\ll1,$ however, Eq. \eqref{eq:ion} takes the form\begin{equation}
\frac{d}{dk}\left[\frac{2(1-\hat{\omega}^{-1})+(1+\eta_{i}+\tau)k^{2}}{(1+\eta_{i}+\tau)k^{2}}\right]\frac{d\hat{J}}{dk}+\left(\frac{1}{4}-\mu^{2}\right)\frac{\hat{J}}{k^{2}}=0.\end{equation}
This has a solution \begin{equation}
\begin{split} & \hat{J}(k)=a_{1}\,_{2}F_{1}\left(-\frac{1}{4}+\frac{\mu}{2},-\frac{1}{4}-\frac{\mu}{2};-\frac{1}{2};y\right)\\
 & +a_{2}y^{3/2}\,_{2}F_{1}\left(\frac{5}{4}+\frac{\mu}{2},\frac{5}{4}-\frac{\mu}{2};\frac{5}{2};y\right),\end{split}
\label{eq:ionhighbeta}\end{equation}
where $_{2}F_{1}$ is the hypergeometric function \citep{abramhyper},
\begin{equation}
y=-\kappa^{2}k^{2},\,\,\,\,\,\,\,\kappa^{2}=\frac{1+\eta_{i}+\tau}{2(1-\hat{\omega}^{-1})},\end{equation}
and $a_{1}$ and $a_{2}$ are two constants to be determined by matching
solution \eqref{eq:ionhighbeta} to the ideal MHD boundary condition
\eqref{eq:bcion} for $k\ll1.$ This matching condition yields \begin{equation}
\frac{a_{2}}{a_{1}}=i\frac{\pi}{3\kappa}\frac{\hat{\beta}(1+\tau+\eta_{i})}{\Delta^{\prime}\rho_{i}}.\end{equation}
The large $k$ behaviour of solution \eqref{eq:ionhighbeta} is given
by \begin{equation}
\begin{split} & \hat{J}(k)\sim y^{1/4-\mu/2}\left[a_{1}\frac{\Gamma(-1/2)\Gamma(-\mu)}{\Gamma^{2}(-1/4-\mu/2)}-ia_{2}\frac{\Gamma(5/2)\Gamma(-\mu)}{\Gamma^{2}(5/4-\mu/2)}\right]\\
 & +y^{1/4+\mu/2}\left[a_{1}\frac{\Gamma(-1/2)\Gamma(\mu)}{\Gamma^{2}(-1/4+\mu/2)}-ia_{2}\frac{\Gamma(5/2)\Gamma(\mu)}{\Gamma^{2}(5/4+\mu/2)}\right]\,\,\,\mbox{for}\,\,\, k\rightarrow\infty,\end{split}
\label{eq:asion}\end{equation} where $\Gamma$ is the Euler Gamma function \citep{abramgamma}.
Thus, the asymptotic form given by Eq. \eqref{eq:asion} implies the
quantites $\hat{a}_{\pm}$ in Eq. \eqref{eq:largkion} are related
by\begin{equation}
\frac{\hat{a}_{-}}{\hat{a}_{+}}=\kappa^{-2\mu}\frac{\Gamma(-\mu)}{\Gamma(\mu)}\frac{\frac{1}{\Gamma^{2}(-\mu/2-1/4)}+i\frac{3}{8}\frac{a_{2}}{a_{1}}\frac{1}{\Gamma^{2}(-\mu/2+5/4)}}{\frac{1}{\Gamma^{2}(\mu/2-1/4)}+i\frac{3}{8}\frac{a_{2}}{a_{1}}\frac{1}{\Gamma^{2}(\mu/2+5/4)}}.\end{equation}
In region $2$ we follow the treatment of Ref. \citep{drake:2509}
to solve Eq. \eqref{eq:elecurr}, obtaining a solution that is small
at $s=0:$\begin{equation}
J(s)=\frac{\sigma_{0}+\sigma_{1}s^{2}}{1+\bar{d}_{0}s^{2}+\bar{d}_{1}s^{4}}\sqrt{\frac{s}{s_{t}}}K_{\mu}\left(\frac{s_{t}}{s}\right),\label{eq:drakehigh}\end{equation}
where $K_{\mu}$ is a modified Bessel function \citep{abrambessel} and
$s_{t}^{2}=1.71\eta_{e}\hat{\beta}/d_{1}.$ For large $s$ this has
the form\begin{equation}
J(s)\sim\frac{1}{\Gamma(1-\mu)}\left(\frac{2s}{s_{t}}\right)^{-3/2+\mu}-\frac{1}{\Gamma(1+\mu)}\left(\frac{2s}{s_{t}}\right)^{-3/2-\mu},\,\,\, s\rightarrow\infty.\label{eq:largesel}\end{equation}
Calculating the Fourier transform of expression \eqref{eq:largesel}
we obtain\begin{equation}
\frac{b_{-}}{b_{+}}=-\left(\frac{s_{t}}{2}\right)^{2\mu}\frac{\Gamma(1-\mu)}{\Gamma(1+\mu)}.\end{equation}
Thus, with the aid of Eq. \eqref{eq:rule}, we have\begin{equation}
\frac{\hat{c}_{-}}{\hat{c}_{+}}=-\left(\frac{2}{s_{t}}\right)^{2\mu}\frac{\Gamma(\mu-1/2)\Gamma(1+\mu)}{\Gamma(-\mu-1/2)\Gamma(1-\mu)}\tan\left[\frac{\pi}{2}\left(\frac{1}{2}+\mu\right)\right].\end{equation}
Finally, matching region $1$ and region $2$ solutions, we obtain
the dispersion relation\begin{equation}
e^{i\frac{\pi}{2}\mu}R^{\mu}=\frac{(\mu+1/2)\Gamma^{2}(-\mu)}{(-\mu+1/2)\Gamma^{2}(\mu)}\left\{ \frac{\mathcal{D}-\cot\left[\pi\left(\frac{1}{4}+\frac{\mu}{2}\right)\right]}{\mathcal{D}-\cot\left[\pi\left(\frac{1}{4}-\frac{\mu}{2}\right)\right]}\right\} ,\label{eq:disprelhighbeta}\end{equation}
where\begin{equation}
R=\frac{8d_{1}}{1.71\eta_{e}}\frac{\rho_{i}^{2}}{\delta_{0}^{2}}\frac{1+\tau+\eta_{i}}{1/4-\mu^{2}},\end{equation}
and \begin{equation}
\mathcal{D}=\frac{\sqrt{2}}{\pi}\Delta^{\prime}\rho_{i}\frac{(1+\tau+\eta_{i})^{1/2}}{(\hat{\omega}-1)^{1/2}}\frac{\Gamma(5/4-\mu/2)\Gamma(5/4+\mu/2)}{\Gamma(3/4-\mu/2)\Gamma(3/4+\mu/2)}.\end{equation}
Introducing the collisionality parameter $C$ of Ref. \citep{drake:2509},\begin{equation}
C=0.51\frac{\nu_{e}}{\omega_{*e}}\frac{m_{e}}{m_{i}}\frac{L_{s}^{2}}{L_{n}^{2}}=2.04\frac{\delta_{0}^{2}}{\tau\rho_{i}^{2}},\label{eq:coll}\end{equation}
equation \eqref{eq:disprelhighbeta} can be recognised as essentially
identical to their semi-collisional, high $\hat{\beta},$ cold ion
result, generalised to finite $T_{i}/T_{e}$ and $\eta_{i}.$ While
surprising at first sight, we see that the solution \eqref{eq:ionhighbeta}
of Eq. \eqref{eq:ion} only utilised the low $k$ expansion of $F(k),$
corresponding to a small ion Larmor radius approximation, although
the solution remained valid for finite $k.$ Furthermore, this is
the solution that we would have obtained using a Pad\'e approximation
to $F(k),$ as in Ref. \citep{pegoraro:364}; however, here it is
exact for $\hat{\beta}\gg1.$ As is evident from our low $\hat{\beta}$
result, this approximation fails to capture the {}``$1/k$ tail''
of $F(k)$ at high $k$ in Eq. \eqref{eq:asF} that is essential to
the result of Ref. \citep{cowley:3230}. 

As shown in Ref. \citep{drake:2509}, the dispersion relation \eqref{eq:disprelhighbeta}
predicts stability for all values of $\Delta^{\prime}.$ An analysis
at large $\Delta^{\prime}$ shows the damping rate becomes small\begin{equation}
\hat{\gamma}\sim-1/(R^{1/2}\Delta^{\prime}\rho_{i}).\end{equation}
The result \eqref{eq:growthKAW}, representing the KAW branch with
$\hat{\omega}\approx1$ at lower $\hat{\beta},$ also has a growth
rate that tends to zero as $(\Delta^{\prime}\rho_{i}\hat{\beta}^{3})^{-1}.$
In the next Section we present a model that demonstrates how the transition
to stability occurs at finite $\hat{\beta}.$

\section{Transition to Stability at Finite $\hat{\beta}$\label{sec:finitebeta}}

As we have seen, the KAW branch has a frequency approaching $\hat{\omega}\approx1$
as $\hat{\beta}$ increases, as also does the high $\hat{\beta}$
theory, and we exploit this property to explore the effect of finite
$\hat{\beta},$ i.e. $\hat\beta\sim\mathcal O (1).$ For finite $s,$ Eq. \eqref{eq:elecurr} takes the
form\begin{equation}
\frac{d^{2}}{ds^{2}}\left\{ 1+[d_{0}-1.71\eta_{e}/(1+\tau)]s^{2}+d_{1}s^{4}\right\} J-1.71\eta_{e}\hat{\beta}J=0.\label{eq:elspecial}\end{equation}
If we consider the special case\begin{equation}
d_{0}-1.71\eta_{e}/(1+\tau)=2\sqrt{d_{1}},\label{eq:special}\end{equation}
corresponding to $\eta_{e}=2.53$ for $\tau=1,$ and introduce $u=d_{1}^{1/4}s,$
then we can write Eq. \eqref{eq:elspecial} as\begin{equation}
\frac{d^{2}}{du^{2}}\left(1+u^{2}\right)^{2}J-\lambda^{2}J=0,\,\,\,\,\lambda^{2}=\frac{1.71\eta_{e}\hat{\beta}}{\sqrt{d_{1}}}=\sqrt{d_{1}}s_{t}^{2}.\end{equation}
The solution, even at $s=0,$ is\begin{equation}
J(s)=\left(1+u^{2}\right)^{-3/2}\cos\left(\sqrt{1-\lambda^{2}}\arctan u\right),\end{equation}
which has the large $s$ asymptotic behaviour\begin{equation}
J(s)\sim s^{-3}-\frac{\sqrt{1-\lambda^{2}}}{d_{1}^{1/4}}\tanh\left(\frac{\pi}{2}\sqrt{\lambda^{2}-1}\right)s^{-4},\,\,\, s\gg1.\label{eq:mia}\end{equation}
When $s$ is such that $(\hat{\omega}-1)s^{2}\sim1,$ the solution
for $J(s)$ is obtained in terms of modified Bessel functions, as in
Eq. \eqref{eq:drakehigh}, except that now we no longer impose the
vanishing boundary condition for small $s,$ but match to the form
\eqref{eq:mia} instead. Thus we obtain\begin{equation}
J(s)\propto\frac{\lambda^{2}/\hat{\beta}\sqrt{d_{1}}-(\hat{\omega}-1)s^{2}}{s^{7/2}}\left[\tilde{a}_{1}K_{\mu}\left(\frac{s_{t}}{s}\right)+\tilde{a}_{2}I_{\mu}\left(\frac{s_{t}}{s}\right)\right],\label{eq:miamia}\end{equation}
where $\tilde{a}_{1}$and $\tilde{a}_{2}$ are constants and we note
$\mu^{2}=1/4-(\hat{\omega}-1)\hat{\beta}\approx1/4.$ Matching the
small argument limit of this solution to the form \eqref{eq:mia},
with $\mu\approx1/2$ and $\hat{\omega}\approx1,$ yields the relationship
between $\tilde{a}_{1}$and $\tilde{a}_{2}:$\begin{equation}
\frac{\tilde{a}_{2}}{\tilde{a}_{1}}=\frac{\pi}{2}\left[1-\frac{\sqrt{\lambda^{2}-1}}{\lambda}\tanh\left(\frac{\pi}{2}\sqrt{\lambda^{2}-1}\right)\right].\end{equation}
At high $\hat{\beta},$ i.e. high $\lambda,$ we see that $\tilde{a}_{2}/\tilde{a}_{1}\rightarrow0,$
and we recover the results in the previous Section.

When $(\hat{\omega}-1)s^{2}\gg1,$ the solution \eqref{eq:miamia}
has the correct form to match to the solution \eqref{eq:asion} in
region $1.$ As a result we obtain a dispersion relation \begin{equation}
e^{i\frac{\pi}{2}\mu}R^{\mu}=\Lambda\left(\mu,\lambda\right)\frac{(\mu+1/2)\Gamma^{2}(-\mu)}{(-\mu+1/2)\Gamma^{2}(\mu)}\left\{ \frac{\mathcal{D}-\cot\left[\frac{\pi}{2}\left(\frac{1}{2}+\mu\right)\right]}{\mathcal{D}-\cot\left[\frac{\pi}{2}\left(\frac{1}{2}-\mu\right)\right]}\right\} ,\label{eq:disptearingfbeta}\end{equation}
which only differs from expression \eqref{eq:disprelhighbeta} by
the key {}``shielding factor'' $\Lambda\left(\mu,\lambda\right):$\begin{equation}
\Lambda\left(\mu,\lambda\right)=1-\sin(\pi\mu)\left[1-\frac{\sqrt{\lambda^{2}-1}}{\lambda}\tanh\left(\frac{\pi}{2}\sqrt{\lambda^{2}-1}\right)\right].\end{equation}
For consistency we need to consider $\Lambda\left(\mu,\lambda\right)$
in the limit $\mu\rightarrow1/2.$ We note that then $\Lambda\approx1$
for $\lambda\gg1,$ $\Lambda=0$ for $\lambda=1$ and $\Lambda\approx-4/(\pi\lambda^{3})$
for $\lambda\ll1.$

Recalling $R\gg1,$ we observe that Eq. \eqref{eq:disptearingfbeta}
has a solution with $\mu\approx1/2,$ as required if our ordering
is to be consistent. This solution is obtained by balancing the large
left-hand-side of Eq. \eqref{eq:disptearingfbeta} by approaching
the zero of the denominator on its right-hand-side. Writing $\mu\approx1/2-\hat{\beta}\delta\hat{\omega},$
we find\begin{equation}
\hat{\omega}\approx1+\frac{2}{1+\tau+\eta_{i}}\left(\frac{2}{\hat{\beta}\Delta^{\prime}\rho_{i}}\right)^{2}+\frac{2\sqrt{2}\Lambda}{\hat{\beta}^{3/2}\Delta^{\prime}\rho_{i}(1+\tau+\eta_{i})}\left(\frac{1.71}{d_{1}}\eta_{e}\right)^{1/2}\frac{\delta_{0}}{\rho_{i}},\label{eq:realmia}\end{equation}
\begin{equation}
\hat{\gamma}\approx-4\sqrt{2}\Lambda\frac{\delta_{0}}{\rho_{i}}\left(\frac{1.71}{d_{1}}\frac{\eta_{e}}{\hat{\beta}}\right)^{1/2}\frac{1}{\hat{\beta}\Delta^{\prime}\rho_{i}(1+\tau+\eta_{i})}\left(1+\Lambda\frac{\Delta^{\prime}\rho_{i}\hat{\beta}^{1/2}}{2\sqrt{2}}\frac{\delta_{0}}{\rho_{i}}\sqrt{\frac{1.71}{d_{1}}\eta_{e}\hat{\beta}}\right),\label{eq:growthmia}\end{equation}
which shows that we require $\hat{\beta}\Delta^{\prime}\rho_{i}\gg1$
for consistency. In the small $\hat{\beta}$ limit the growth rate
reduces to\begin{equation}
\hat{\gamma}\approx\frac{32}{\pi\sqrt{2}}\frac{\delta_{0}}{\rho_{i}}\frac{d_{1}^{1/4}}{1.71\eta_{e}}\frac{1}{\hat{\beta}^{3}\Delta^{\prime}\rho_{i}(1+\tau+\eta_{i})}\left(1-\frac{\sqrt{2}d_{1}^{1/4}}{1.71}\frac{\Delta^{\prime}\rho_{i}}{\eta_{e}\hat{\beta}}\frac{\delta_{0}}{\rho_{i}}\right).\label{eq:growthmialim}\end{equation}
Taking account of the condition \eqref{eq:special} on $\eta_{e},$
we see that this coincides with the low $\hat{\beta}$ result \eqref{eq:growthKAW}
of the KAW; similarly, the frequency $\hat{\omega}$ reduces to the
result \eqref{eq:realKAW}. 

Equation \eqref{eq:growthmia} implies stability if $\Lambda>0.$ This
stability boundary corresponds to $\lambda=1,$ so that there is stability
for values of $\hat{\beta}$ satisfying \begin{equation}
\hat{\beta}>\hat\beta_c\equiv\frac{\sqrt{d_{1}}}{1.71\eta_e}=0.34,\label{eq:stabmia}\end{equation} for $\tau=1.$
Thus the model correctly predicts stability at high $\hat{\beta}$
as found in Section \ref{sec:highbeta}. If $\Lambda<0$ then there
is still stability if $\Delta^{\prime}$ is sufficiently large:\begin{equation}
\Delta^{\prime}\delta_{0}\sqrt{\frac{1.71}{d_{1}}\hat{\beta}\eta_{e}}>-\frac{1}{\Lambda}.\label{eq:last}\end{equation} For $\hat\beta\rightarrow\hat\beta_c,$ one finds stability if \begin{equation}
\Delta^{\prime}\delta_{0}>\frac{\pi}{4\sqrt 2}\sqrt{\frac{1.71\eta_e}{d_1}}\frac{1}{\hat\beta_c-\hat\beta}.\label{eq:ecche}\end{equation} 
As $\hat{\beta}$ reduces, $-\Lambda$ increases and the necessary
value of $\Delta^{\prime}$ for stability decreases. At low $\hat{\beta}$
the result is consistent with the stability condition \eqref{eq:stabKAW}.
We note that a similar analysis
can be performed for another special case:\begin{equation}
d_{0}-1.71\eta_{e}/(1+\tau)=-2\sqrt{d_{1}}.\label{eq:nrespec}\end{equation}
Now, however, we find the analogue of the screening factor $\Lambda$
is always positive so the mode is always stable. Condition \eqref{eq:nrespec},
which guarantees stability at finite $\hat{\beta},$ coincides with
condition \eqref{eq:special}, the stability condition at low $\hat{\beta},$
so we have a consistent picture.%

Although our model equation is strictly valid for a specific, though
reasonable, value of $\eta_{e},$ it captures in a clear way the onset
of screening by plasma gradients that stabilises modes in high $\hat{\beta}$
plasma.

\section{Tearing Mode stability Diagram}\label{sec:stabdiagr}

\begin{figure}[h]
\includegraphics[height=0.3\textheight]{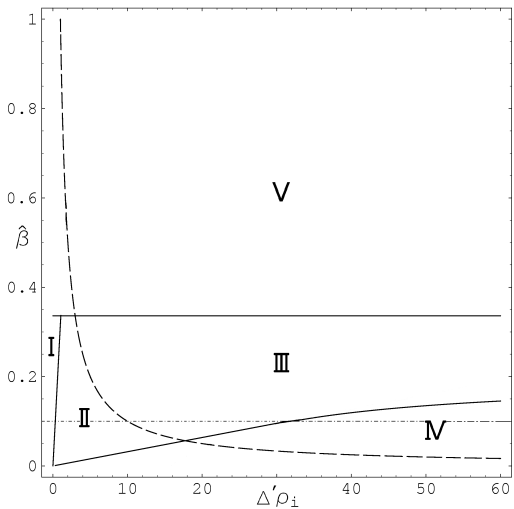} 
\caption{Stability diagram for the drift-tearing/kinetic Alfv\'en mode in terms of $\Delta^{\prime}\rho_{i}$
and $\hat{\beta}.$ Regions I, IV and V are stable, Regions II and III unstable. The finite $\hat{\beta}$
analysis is not valid in the region below the hyperbola $\hat \beta \Delta^{\prime}\rho_i=1$ (dashed line); the low $\hat\beta$ theory fails below the dotted line $\hat\beta=(\delta_0/\rho_i)^{1/2}$ when $\hat\beta\Delta^{\prime}\rho_i\sim\mathcal O(1).$ Here $\delta_0/\rho_i=10^{-2}.$}
\label{fig:fig7}
\end{figure}

In this section we draw together results from previous sections concerning the stability of the 
drift-tearing and kinetic Alfv\'en waves, summarising them in Fig. \ref{fig:fig7}, a diagram in the space of $\hat \beta$
and $\Delta^{\prime}\rho_i$ which delineates stable and unstable regions and regions of validity of these results.

The low $\hat\beta$ theory of Section \ref{subsec:lowbetatearing}, valid for $\hat\beta\ll1,$ shows that the drift-tearing
mode is stable in Region I, becoming unstable in Region II when $\Delta^{\prime}\rho_i$ exceeds the critical
value given in criterion \eqref{eq:critdelprime}. When the approximate boundary $\Delta^{\prime}\rho_i\hat\beta\sim\mathcal{O}(1)$
is exceeded as in Region III, the analysis in Section \ref{subsec:KAW} is valid and the kinetic Alfv\'en
wave becomes the unstable branch; this remains unstable until the criterion \eqref{eq:stabKAW} is satisfied and one enters the stable Region IV. 
Above a critical value of $\hat\beta=\hat\beta_{c}\sim\mathcal O(1),$ given precisely by condition \eqref{eq:stabmia} in Section \ref{sec:finitebeta} for the 
special case of $\eta_e$ specified in Eq.\eqref{eq:special}, the kinetic Alfv\'en wave is stable, as represented by Region V. 
This stability region extends to high $\hat\beta$ as shown in Section IV. The boundary between Regions IV and III asymptotes to the critical value of $\hat\beta$ following the scaling $\Delta^{\prime}\propto(\hat\beta_c-\hat\beta)^{-1}.$ The low $\hat\beta$ results discussed above are limited by the 
condition $\hat\beta>{(\delta_0/\rho_i)}^{1/2}$ (dotted line in Fig. \ref{fig:fig7}) when $\Delta^{\prime}\rho_i\hat\beta\sim\mathcal O (1).$ Below the dotted line, there is still an instability when $\Delta^{\prime}\rho_i$
exceeds the critical value from the criterion \eqref{eq:critdelprime}, beyond which the growth rate continues to increase with $\Delta^{\prime}\rho_i,$
eventually merging with the dissipative kink mode, discussed in the next Section, when $\Delta^{\prime}\rho_i\gg1.$

\section{The Internal Kink Mode\label{sec:The-Internal-Kink}}

\subsection{Low $\hat{\beta}$ theory}

In this Section we analyse the internal kink mode. When $\Delta^{\prime}<0$
we first follow Pegoraro et al. \citep{pegoraro:364} in introducing
the growth rate of the ideal MHD internal kink mode: $\lambda_{H}=-\pi/\Delta^{\prime}r_{s},$
where $q(r_{s})=1.$ The solution of the branch $C(\hat{\omega})=0$
of the general dispersion relation \eqref{eq:dispJim} then reproduces
the FLR modified internal kink mode dispersion relation obtained,
and analysed numerically, in Ref. \citep{pegoraro:364}. The small
terms on the left-hand-side of Eq. \eqref{eq:dispJim} can be retained
to obtain a correction to the ideal result. However, in the limit
$\lambda_{H}\rightarrow0$ these terms must be retained to obtain
the dispersion relation for the dissipative internal kink mode which
is associated with the $\hat{\omega}\approx0$ branch of Eq. \eqref{eq:solfund}.

\begin{figure}[h]
\includegraphics[height=0.26\textheight]{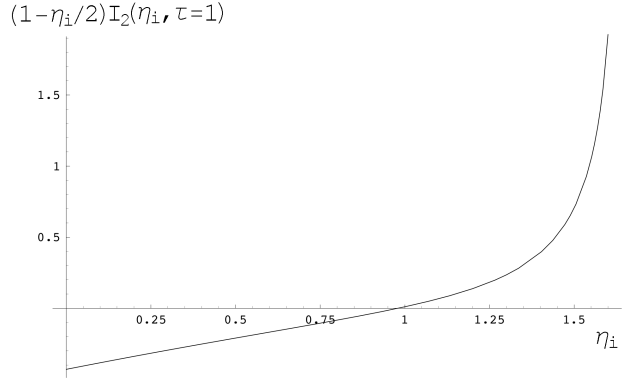} 

\caption{The function $(1-\eta_{i}/2)I_{2}$ appearing in Eq. \eqref{eq:i2lowbetakink}
as a function of $\eta_{i}$ for $\tau=1$.}
\label{fig:fig8}
\end{figure}

For small values of $|\hat{\omega}|$ and $\hat{\beta}\ll1,$ we can
approximate the terms in Eq. \eqref{eq:dispJim}:\begin{equation}
\begin{split} & A(\hat{\omega})=\sqrt{1+\tau}\sqrt{d_{0}-(1+1.71\eta_{e})}\equiv\sqrt{(1+\tau)d(\eta_{e})}\\
 & B(\hat{\omega})=\frac{\Delta^{\prime}\rho_{i}}{\pi\hat{\beta}}\hat{\omega}\\
 & C(\hat{\omega})=1+\frac{\Delta^{\prime}\rho_{i}}{\pi}\hat{\beta}\hat\omega I(\hat{\omega},\tau,\eta_{i})\\
 & D(\hat{\omega})=-\sqrt{d_{1}},\end{split}
\label{eq:kinkcoeff}\end{equation} where now we require $\Delta^{\prime}\rho_i|\hat \omega|/(\pi\hat\beta)\gg 1.$
Assuming $\hat{\omega}$ to be real and negative, i.e. in the ion
direction, we can calculate the integrals analytically:\begin{equation}
I(\hat{\omega},\eta_{i})=-\sqrt{\pi}I_{2}(\eta_{i},\tau)-\frac{\sqrt{\pi}}{1-\eta_{i}/2}\ln\left[\frac{1-\eta_{i}/2}{-\hat{\omega}(1+\tau)\sqrt{\pi}}\right],\label{eq:kink1}\end{equation}
where\begin{equation}
I_{2}(\eta_{i},\tau)=\frac{1}{\sqrt{\pi}}\int_{0}^{\infty}\frac{dk}{k^{2}}\left[\frac{F_{0}}{G_{0}}-\frac{\sqrt{\pi}}{1-\eta_{i}/2}\frac{k^{2}}{(1+k)}\right].\label{eq:i2lowbetakink}\end{equation}
The quantity $(1-\eta_{i}/2)I_{2}(\eta_{i},\tau)$ is plotted in Fig.
\ref{fig:fig8} as a function of $\eta_{i}$ for $\tau=1.$%
For consistency, the present analysis of Eq. \eqref{eq:dispJim} is
limited to the situation where $\delta_{0}/\rho_{i}\hat{\beta}\ll1$
to ensure $|\hat{\omega}|\ll1.$

Substituting the result \eqref{eq:kinkcoeff} into Eq. \eqref{eq:KAWdisp}
and expressing $\Delta^{\prime}$ in terms of $\lambda_{H},$ we obtain
a dispersion relation\begin{equation}
\begin{split} & \hat{\omega}\left\{ \ln\left[\frac{(1+\tau)\sqrt{\pi}}{1-\eta_{i}/2}\hat{\omega}\right]-i\pi-\left(1-\frac{\eta_{i}}{2}\right)I_{2}-\frac{1}{\sqrt{\pi}}\left(1-\frac{\eta_{i}}{2}\right)\frac{\lambda_{H}}{\rho_{i}}\frac{1}{\hat{\beta}\hat{\omega}}\right\} =\\
 & \sqrt{\frac{1+\tau}{\pi d_{1}}d(\eta_{e})}\left(1-\frac{\eta_{i}}{2}\right)\frac{\delta_{0}}{\rho_{i}\hat{\beta}^{2}}e^{-i\frac{\pi}{4}},\end{split}
\label{eq:newdispkink}\end{equation} which is valid for $|\lambda_H/\rho_i|\ll 1.$
For the special case $\lambda_{H}=0$ this produces, as in Ref. \citep{0029-5515-33-3-I01},
an unstable mode rotating in the ion direction\begin{equation}
\hat{\omega}=-\sqrt{d(\eta_{e})\frac{1+\tau}{2\pi d_{1}}}\left(1-\frac{\eta_{i}}{2}\right)\frac{1}{\ln(\rho_{i}\hat{\beta}^{2}/\delta_{0})}\frac{\delta_{0}}{\rho_{i}\hat{\beta}^{2}}e^{-i\frac{\pi}{4}},\label{eq:ionkink}\end{equation}
where we have iterated on the $\ln(-\hat{\omega}^{-1})$ term, ignoring
$\mathcal{O}(1)$ corrections. When $\delta_{0}/\rho_{i}\ll\hat{\beta}\left|\lambda_{H}\right|/\rho_{i}\ll\hat{\beta}^{2}$
we obtain a stable mode in the electron direction:\begin{equation}
\hat{\omega}=\frac{1}{\sqrt{\pi}}\left(1-\frac{\eta_{i}}{2}\right)\frac{\left|\lambda_{H}\right|}{\rho_{i}\hat{\beta}\,\ln(\rho_{i}\hat{\beta}/\left|\lambda_{H}\right|)}\left[1-\frac{i\pi}{\ln(\rho_{i}\hat{\beta}/\left|\lambda_{H}\right|)}\right].\label{eq:electronkink}\end{equation}
Thus there is a marginally stable value of $\lambda_{H},$ which can
be found by equating real and imaginary parts, with $\hat{\omega}$
real, in Eq. \eqref{eq:newdispkink}, to yield\begin{equation}
\lambda_{H}^{crit}=-\delta_{0}\frac{1}{\pi\hat{\beta}}\sqrt{d(\eta_{e})\frac{1+\tau}{2d_{1}}}\left\{ \ln\left[\hat{\beta}^{2}\frac{\rho_{i}}{\delta_{0}}\frac{\pi}{(1+\tau)^{3/2}}\sqrt{\frac{2d_{1}}{d(\eta_{e})}}\right]+\pi+\left(1-\frac{\eta_{i}}{2}\right)I_{2}(\eta_{i})\right\} .\label{eq:critlambdaH}\end{equation}
The condition that $|\hat{\omega}|$ remains small at these value of
$\lambda_{H}$ requires \begin{equation}
\hat{\beta}>\left(\frac{\delta_{0}}{\rho_{i}}\right)^{1/2}.\end{equation}
We note that the numerical solution at high $\Delta^{\prime}\rho_i$ in the region
$\delta_0/\rho_i\gg\hat\beta^2$ described in Section \ref{subsec:KAW} can be followed
through the boundary $\hat\beta\sim{(\delta_0/\rho_i)}^{1/2}$ to merge with the result \eqref{eq:ionkink}.
We identify also a strong stabilising effect from $\eta_{e}:$
Eq. \eqref{eq:kinkcoeff} implies the factor $d(\eta_{e})$ becomes
negative, corresponding to stability, if\begin{equation}
\eta_{e}>(d_{0}-1)/1.71\approx2.38.\end{equation}
Using Eq. \eqref{eq:coll} we can express the results \eqref{eq:ionkink},
\eqref{eq:electronkink} and \eqref{eq:critlambdaH} in terms of the
collisionality parameter $C$ of Ref. \citep{drake:2509}.

\subsection{Finite $\hat{\beta}$ theory}

Because the dissipative internal kink mode corresponds to $|\hat{\omega}|\ll1$
we can seek a solution of Eqs. \eqref{eq:ion} and \eqref{eq:elecurr}
based on an expansion in $\hat{\omega}$ at finite $\hat{\beta},$
rather that in $\hat{\beta}$ at finite $\hat{\omega},$ as in Section
\ref{sec:lowbetaJIM}. The details are presented in the Appendix but
we note the essential points here. For the ions we can exploit the
same expansion as in Eq. \eqref{eq:pegsolions}, but now based on
$|\hat{\omega}|\ll1;$ in this region of $k$, $F(k)$ and $G(k)$ take
simpler forms. However, this expansion fails at large $k\sim\hat{\omega}^{-1}$
when the finite $\hat{\omega}$ correction to $G(k)$ becomes important.
The ion response must be calculated in this intermediate region with
$k\sim\hat{\omega}^{-1}$ before the solution can be matched to the
electron region. The electron solution, i.e. the solution of Eq. \eqref{eq:elecurr}
calculated in real space, also exploits the condition $|\hat{\omega}|\ll1$
by considering two sub-regions. In the first of these we consider
$s\sim\mathcal{O}(1)$ when we take advantage of the fact that the
quadratic term in the demonimator of the conductivity in Eq. \eqref{eq:electr}
can be ignored. In the second sub-region we take $s\sim|\hat{\omega}|^{-1/2}\gg1$
which again simplifies the equation so that an analytic solution can
be obtained. We match the solutions in the two sub-regions at intermediate
values of $s$, deduce the power-like asymptotic form for $s\gg|\hat{\omega}|^{-1/2}$
and then use Eq. \eqref{eq:rule} to obtain the solution in $k-$space.
Finally we match this electron region solution to the intermediate
ion region solution to provide a dispersion relation for the dissipative
internal kink mode. This relation is given by Eq. \eqref{eq:dispkinkarbbetaapp}
in the Appendix, which we reproduce here:\begin{equation}
\left[\frac{e^{-i\frac{5}{4}\pi}}{8\sqrt{\pi}\hat{\omega}}\frac{\delta_{0}}{\rho_{i}}\sqrt{\frac{d(\eta_{e})}{d_{1}}}\,\frac{1-\eta_{i}/2}{(1+\tau)^{3/2}}\right]^{2\mu_{1}}=\frac{1/2-\mu_{1}}{1/2+\mu_{1}}\frac{\Gamma^{2}(\mu_{1})}{\Gamma^{2}(-\mu_{1})}\frac{\Sigma(\hat{\omega})+i\frac{\hat{\beta}}{1+\tau}e^{i\pi\mu_{1}}}{\Sigma(\hat{\omega})+i\frac{\hat{\beta}}{1+\tau}e^{-i\pi\mu_{1}}}.\label{eq:dispkinkarbbeta}\end{equation}
We recall that we can express $\delta_{0}/\rho_{i}$ in terms of the
collisionality parameter $C,$ {[}note $\delta_{0}/(8\rho_{i})=0.088\tau^{1/2}C^{1/2}$].
Here $\Sigma$ is

\begin{equation}
\begin{split} & \Sigma(\hat{\omega})=-\frac{\cos[\pi\mu_{1}]}{\pi}\left\{ 1-\frac{1}{\sqrt{\pi}}\frac{1-\eta_{i}/2}{1+\tau}\frac{\lambda_{H}}{\rho_{i}}\frac{1}{\hat{\omega}}-\hat{\beta}\frac{1-\eta_{i}/2}{1+\tau}I_{2}\right.\\
 & \left.+\frac{\hat{\beta}}{1+\tau}\left[\frac{3}{2}-k_{0}+\ln\left(\frac{\sqrt{\pi}(1+\tau)}{1-\eta_{i}/2}\hat{\omega}\right)\right]\right\} ,\end{split}
\label{eq:sigma}\end{equation}
with $2\mu_{1}=\sqrt{1+4\hat{\beta}/(1+\tau)},$ $k_{0}=\psi(1)+\psi(3)-\psi(3/2-\mu_{1})-\psi(3/2+\mu_{1}),$
where $\psi$ is the digamma function \citep{abramgamma}. We can solve
this analytically in the low $\hat{\beta}$ limit. Assuming the logarithmic
term in Eq. \eqref{eq:sigma} dominates, we recover the result \eqref{eq:newdispkink}.
However Eq. \eqref{eq:dispkinkarbbeta} allows us to explore the effect
of finite $\hat{\beta}$ on the stability of the dissipative internal
kink mode. %
\begin{figure}[h]
\includegraphics[height=0.3\textheight]{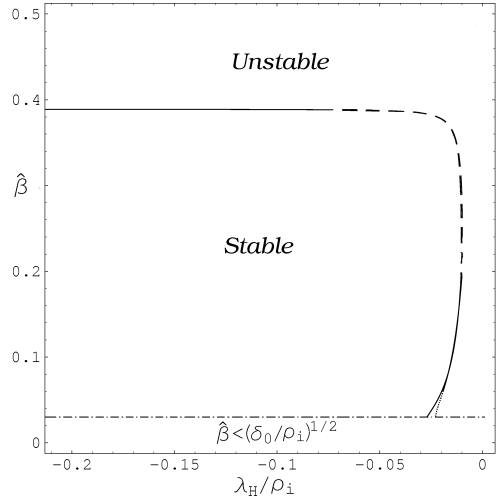} 

\caption{The boundary of marginal stability for the dissipative internal kink
mode when |$\lambda_H/\rho_i|\ll1.$ The complete line is from the numerical solution of the dispersion
relation \eqref{eq:dispkinkarbbeta}. The dashed line is the analytical
boundary of Eq. \eqref{eq:kinkbounfarb}, the dotted line is the low
$\hat{\beta}$ analytical limit of Eq. \eqref{eq:critlambdaH}. Here
$\delta_{0}/\rho_{i}=10^{-3},$ $\eta_{e}=\tau=1,$ $\eta_{i}=0.$
In the region $\hat\beta<(\delta_{0}/\rho_{i})^{1/2},$ the finite
$\hat \beta $ theory breaks down.}
\label{fig:fig9}
\end{figure}
\begin{figure}[h]
\includegraphics[height=0.3\textheight]{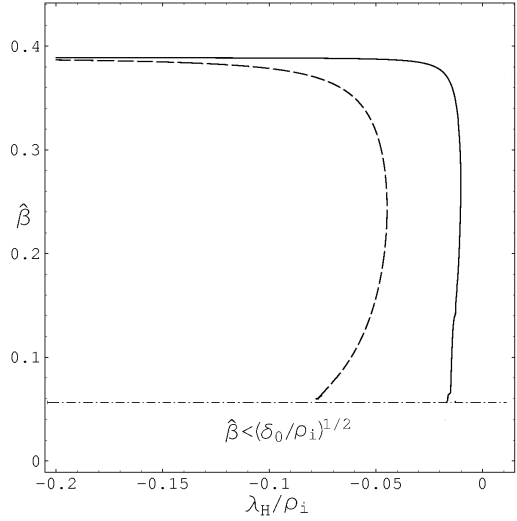} 

\caption{Boundary of the marginal stability region for the dissipative internal
kink mode from the numerical solution of Eq. \eqref{eq:dispkinkarbbeta}.
Solid line: $\delta_{0}/\rho_{1}=10^{-3},$ dashed line: $\delta_{0}/\rho_{1}=10^{-2}.$ }
\label{fig:fig10}
\end{figure}
For low values of $\hat{\beta}/(1+\tau)$ there is a threshold value
for $\left|\lambda_{H}\right|$ for the onset of the instability,
as in Eq. \eqref{eq:critlambdaH}, but this is lost as $\hat{\beta}/(1+\tau)$
increases beyond $0.194.$ It is again
possible to calculate the marginal value of $\left|\lambda_{H}\right|$
by equating real and imaginary parts of Eq. \eqref{eq:dispkinkarbbeta}
with real $\hat{\omega}$ (for a consistent solution we must take
$\hat{\omega}=\exp(-2\pi i)\hat{\omega}_{1},$ with $\hat{\omega}_{1}$
real and positive):\begin{equation}
\begin{split} & \lambda_{H}^{crit}=-\delta_{0}H(\mu_{1})\frac{\sqrt{1+\tau}}{8}\sqrt{\frac{d(\eta_{e})}{d_{1}}}\times\\
 & \left\{ \frac{\hat{\beta}}{1+\tau}\frac{\pi}{\cos(\pi\mu_{1})}\sqrt{\frac{\sin(\pi\mu_{1}/2)}{\sin(3\pi\mu_{1}/2)}}-1\right.\\
 & \left.-\frac{\hat{\beta}}{1+\tau}\left[\frac{3}{2}-k_{0}-\left(1-\frac{\eta_{i}}{2}\right)I_{2}+\ln\left(\frac{\delta_{0}}{\rho_{i}}H(\mu_{1})\frac{\sqrt{1+\tau}}{8}\sqrt{\frac{d(\eta_{e})}{d_{1}}}\right)\right]\right\} ,\end{split}
\label{eq:kinkbounfarb}\end{equation}
while $\hat{\omega}$ is given by \begin{equation}
\hat{\omega}=\frac{1}{8}\frac{\delta_{0}}{\rho_{i}}H(\mu_{1})\sqrt{\frac{d_{1}}{\pi d(\eta_{e})}}\frac{1-\eta_{i}/2}{(1+\tau)^{3/2}}.\end{equation}
Here \begin{equation}
H(\mu_{1})=\left\{ \frac{1/2+\mu_{1}}{1/2-\mu_{1}}\frac{\Gamma^{2}(-\mu_{1})}{\Gamma^{2}(\mu_{1})}\frac{\cos(\pi\mu_{1})}{\cos(\pi\mu_{1}/2)\pm\sqrt{\sin(\pi\mu_{1}/2)\sin(3\pi\mu_{1}/2)}}\right\} ^{\frac{1}{2\mu_{1}}}.\end{equation}
Expression \eqref{eq:kinkbounfarb} correctly reduces to the result
\eqref{eq:critlambdaH} in the low $\hat{\beta}$ limit.

Figure \ref{fig:fig9} shows the resulting critical value of $\left|\lambda_{H}\right|/\rho_{i}$
as a function of $\hat{\beta}$ for $\delta_{0}/\rho_{i}=10^{-3}$
(other parameters are $\eta_{e}=1,$ $\eta_{i}=0$ and $\tau=1$);
we recall the low $\hat{\beta}$ limitation to ensure $|\hat{\omega}|\ll1$
which is marked in the figure. The validity regime for the diagram is $|\lambda_H/\rho_i|\ll1,$ i.e. in the region where the mode is not ideal, but close to the ideal stability boundary $\lambda_H/\rho_i=0$. There is no marginal value of $\left|\lambda_{H}\right|$
when $\hat{\beta}/(1+\tau)>7/36\approx0.194,$ corresponding to $\mu_{1}=2/3$
when Eq. \eqref{eq:kinkbounfarb} implies $|\lambda_{H}|\rightarrow\infty.$
Figure \ref{fig:fig10} shows the same critical value of $\left|\lambda_{H}\right|/\rho_{i}$
as in Fig. \ref{fig:fig9}, but for two different values of $\delta_{0}/\rho_{i}:$
$10^{-2}$ and $10^{-3}.$ The distance of the boundary from the line
at $\lambda_{H}\equiv0$ increases with $\delta_{0}/\rho_{i}.$

\section{Sawtooth Modelling\label{sec:Sawtooth-Modelling}}

Large tokamaks can operate in regimes where the collisional fluid
limit for the drift-tearing mode is no longer justified and the semi-collisional
theory is appropriate. The stability theory of the semi-collisional
drift-tearing and internal kink modes is characterised by the two
key parameters $\hat{\beta},$ and $\delta_{0}/\rho_{i}$ and semi-collisional
effects are strong when $\hat{\beta}\sim\mathcal{O}(1)$ and large
ion orbit effects are relevant when $\delta_{0}/\rho_{i}\ll1.$ As
an example, we consider sawtooth modelling in JET and ITER. For the
$m=1$ mode, assuming density profiles such that $n=n(0)(1-\alpha r^{2}/a^{2})$
with $\alpha\sim1/2,$ we can write\begin{equation}
\delta_{0}=\left(\frac{\alpha\nu_{e}}{\Omega_{e}}\right)^{1/2}\frac{R}{aq^{\prime}(r_{1})}\,\mbox{cm};\,\,\,\hat{\beta}=2\beta_{e}\frac{R^{2}}{a^{2}}\frac{1}{a^{2}\left[q^{\prime}(r_{1})\right]^{2}},\,\,\,\mbox{where}\,\, q(r_{1})=1.\end{equation}
We take the following sets of parameters\begin{equation}
\mbox{JET}:\, n_{e}(0)=10^{20}\,\mbox{m}^{-3},\,\, T_{e}(0)=5\mbox{keV},\,\, B=3\mbox{T},\, R=3\mbox{m},\, a=1\mbox{m}.\end{equation}
\begin{equation}
\mbox{ITER}:\, n_{e}(0)=10^{20}\,\mbox{m}^{-3},\,\, T_{e}(0)=22\mbox{keV},\,\, B=5.3\mbox{T},\, R=6.2\mbox{m},\, a=2\mbox{m}.\end{equation}
From these we derive \begin{equation}
\mbox{JET}:\,\rho_{i}=0.3\,\mbox{cm},\,\,\beta_{e}(0)=2.2\times10^{-2},\,\,\nu_{e}/\Omega_{ce}=3\times10^{-9}.\end{equation}
\begin{equation}
\mbox{ITER}:\,\rho_{i}=0.4\,\mbox{cm},\,\,\beta_{e}(0)=3.1\times10^{-2},\,\,\nu_{e}/\Omega_{ce}=2\times10^{-10},\end{equation}
so that\begin{equation}
\hat{\beta}^{JET}=0.4/\left[aq^{\prime}(r_{1})\right]^{2},\,\,\,\,\hat{\beta}^{ITER}=0.6/\left[aq^{\prime}(r_{1})\right]^{2},\end{equation}
\begin{equation}
\delta_{0}^{JET}=1.5\times10^{-4}/q^{\prime}(r_{1})\,\,\mbox{cm},\,\,\,\,\delta_{0}^{ITER}=0.4\times10^{-4}/q^{\prime}(r_{1})\,\,\mbox{cm}.\end{equation}
Clearly in each case $\hat{\beta}\sim\mathcal{O}(1),$ while $\delta_{0}/\rho_{i}\ll1,$
and the semi-collisional regime is appropriate$^1$ \footnotetext{When the collisionality is low enough to give $\delta_{0}\ll d_{e,}$
with $d_{e}=\rho_{e}/\sqrt{\beta_{e}}$ the electron inertia length, the
collisionless regime is appropriate \citep{antonsen-coppi,cowley:3230}.
}.

\begin{figure}[h]
\includegraphics[height=0.3\textheight]{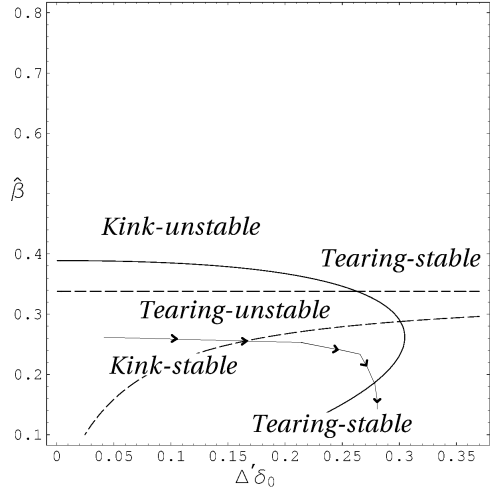} 

\caption{Boundary of marginal stability of the dissipative internal kink mode
and drift-tearing mode, respectively in terms of $\hat\beta$ and $\delta_0\Delta^{\prime}$ from Eq.\eqref{eq:dispkinkarbbeta} (solid line) and from Eq.  \eqref{eq:disptearingfbeta} (dashed lines). The parameters are the same used in Fig. \ref{fig:fig9}.
The kink boundary does not change significantly for $\eta_{e}<2.39.$ A schematic trajectory for the evolution of a discharge following a sawtooth crash is indicated by the line with arrows.}
\label{fig:fig11}
\end{figure}

Superimposing the stability diagrams in Figs. \ref{fig:fig7} and
\ref{fig:fig9} at high $\Delta^{\prime}\rho_i$ (where we replace $-\lambda_{H}$ in terms of $\Delta^{\prime}r_{s}=-\pi/\lambda_{H}$)
as shown in Fig. \ref{fig:fig11}, we see that there are regions of
parameter space %
corresponding to windows of stability to both modes. This has a potential
implication for sawtooth modelling in the spirit of Ref. \citep{0741-3335-38-12-010},
but provides more accurate stability criteria to associate with a trigger
for the sawtooth crash. After such a crash, one expects a relatively
rapid recovery, on the scale of the energy confinement time, of density and temperature, followed by a slower evolution of the $q$ profile
on the resistive diffusion time scale. An expression for $\Delta^{\prime}$ is given
by $r_{1}\Delta^{\prime}=\hat{s}_{1}^{2}/\delta \hat W,$ where $\delta \hat W$
is a measure of the potential energy of the ideal internal kink mode, and $\hat{s}_{1}$
is the magnetic shear at $r_{1},$ the radial position of the $q=1$ surface. $\delta \hat W$ is a rather complicated function of
the pressure and current profiles and, as demonstrated in Ref. \citep{0741-3335-38-12-010},
will also be a sensitive function of the pressure profile of fusion alpha
particles in ITER. However, the contribution of the core plasma, as calculated in Ref. \citep{bussacsaw}
for $r_1/a\ll 1,$ gives $$\delta\hat W \propto {\left(\frac{r_1}{R_0}\right)}^4\left(1-q_0\right)\left(\frac{13}{144}-\beta_p^2\right)$$
when $\beta_p$ in a measure of the poloidal $\beta$ within the $q=1$ surface. As the core plasma heats up, $\delta\hat W$
reduces (due to increasing $\beta_p$) so that, provided that $\hat\beta<0.4$ (see Fig. \ref{fig:fig7}) there
may be a period when the drift-tearing mode is unstable (a possible explanation for successor oscillations \cite{west}
after a sawtooth crash), followed by a stable period during which the equilibrium trajectory in $(\hat\beta,\delta_0\Delta^{\prime})$ (see Fig \ref{fig:fig11}) passes through a region stable to both modes.

After the density and temperature profiles have recovered (we suppose this occurs before meeting the dissipative 
kink-instability boundary on the right of the stable region), the evolution
of $\Delta^{\prime}$ will be determined by that of the $q(r)$ profile, with dependence on both 
$s_1$ (increasing $\Delta^{\prime}$) and $r_1$ (decreasing $\Delta^{\prime};$ however, experimental evidence suggests that the variation of $r_1$ during a sawtooth ramp is weak).
Since $\delta_{0}\propto r_1/\hat{s}_{1},$ the parameter $\delta_{0}\Delta^{\prime}$ may remain fairly constant
during this phase if $r_1$ grows only slowly relatively to $\hat s_1(t)$. However $\hat{\beta}\propto\hat{s}_{1}^{-2}$
should decrease and the equilibrium trajectory
could exit the stable region in $(\hat\beta, \delta_0\Delta^{\prime})$ by crossing
the dissipative kink boundary, as shown schematically in Fig \ref{fig:fig11}.
This could represent the onset of the next sawtooth
collapse phase. The sawtooth trigger can thus be deduced from Eq. \eqref{eq:critlambdaH},
which resembles, but is a more precise formulation of, the stability
criterion invoked by the authors of Ref. \citep{0741-3335-38-12-010}
as a prescription for a sawtooth trigger. 

In summary, the stability diagram in Fig. \ref{fig:fig11} suggests a picture
in which, after a sawtooth crash, there is first a period during which the drift-tearing/kinetic Alfv\'en
wave branches may be unstable, giving rise to postcursor or successor oscillations,
then a phase in which these modes are stabilised and the dissipative internal kink
is not yet unstable, until finally the threshold for the latter mode is reached and a sawtooth
crash ensues: a repetition of this whole cycle then begins.

The discussion above is, of course, extremely qualitative. A more meaningful technique would
be to use a transport code to evolve the density, the temperature,
the alpha particle population if appropriate, and the $q$ profile during a sawtooth
period and follow the resulting discharge trajectory relative to the
stability boundaries in Fig. \ref{fig:fig11}. Here we content ourselves
with simulating the evolution of the $q$ profile following a sawtooth
crash, using this to calculate the evolution of $\hat{\beta}(t).$
The simulations utilise neoclassical resistivity, corrected for finite
$\nu_{*e},$ the banana regime collisionality parameter, since this
becomes important close to the magnetic axis \citep{parkmonticello}.
In the case of JET, $\nu_{*e}=5\times10^{-3},$ while for ITER, $\nu_{*e}=5\times10^{-4}.$
As an example we assume the post-crash $q$ profile is given by the
Kadomtsev full reconnection model \citep{kad2} - other models could
well have been invoked, such as a partial reconnection model \citep{0741-3335-38-12-010}.
We study a typical JET and ITER base-line H-mode scenario. The resulting
prediction for the sawtooth periods, identified as the times that
the instability boundary in Fig. \ref{fig:fig11} are reached as a
result of the drop in $\hat{\beta}(t),$ are quite plausible for both
JET and ITER simulations: $2$ sec for JET, and $200$ sec for ITER.

\section{Discussion and Conclusions\label{sec:Conclusions}}

The stability theory of the semi-collisional drift-tearing and internal
kink modes, the subject of this paper, is characterised by the two
key parameters, $\hat{\beta}$ and $\delta_{0}/\rho_{i}$, the latter
being related to the collisionality parameter, $C$: $\delta_{0}/\rho_{i}\sim C^{1/2}$
. Semi-collisional effects are strong when $\hat{\beta}\sim\mathcal{O}(1)$
and large ion orbit effects are relevant when $\delta_{0}/\rho_{i}\ll1$.
Large, hot tokamaks such as JET and ITER operate in regimes where
these conditions are satisfied. We have derived a unified theory for
the stability of the modes at low $\hat{\beta}=\beta_{e}(L_{s}^{2}/L_{n}^{2})/2$
and for ion Larmor orbits much larger than the semi-collisional layer
width, i.e. $\rho_{i}/\delta_{0}\gg1$, also extending the earlier
work \citep{pegoraro:364} on the kink mode. The effect of finite
$\hat{\beta}$ on the stability of the drift-tearing and dissipative
internal kink modes has then been elucidated and forms the basis for
a sawtooth model. 

The non-local aspects resulting from the ion orbits are accommodated
by working in Fourier transform space, where the problem reduces to
solving a fourth order ordinary differential equation. The problem
is further simplified by exploiting the two disparate scales, $\rho_{i}$
and $\delta_{0}$, solving separately in the \textquoteleft{}ion region\textquoteright{}
where the dissipative electron effects can be ignored and the \textquoteleft{}electron
region\textquoteright{} where the ions can be regarded as un-magnetised.
The ion region is then described by a second order differential equation
in \textquoteleft{}$k$-space\textquoteright{}, with the ideal MHD
solution involving $\Delta^{\prime}$ providing a boundary condition
at low $k$. The high-$k$ power solutions are then asymptotically
matched to the similar ones emanating from the electron region solution.
However, the electron region is described by a fourth order differential
equation in $k$-space and it is easier to solve this by first transforming
back to $x$-space, where it becomes second order, and then calculating
the Fourier transform of the small $x$ asymptotic form of its solution
for matching to the ion region.

The outcome at low $\hat{\beta}$ is a unified dispersion relation
that encompasses the drift-tearing mode and internal kink mode. The
nature of the resulting modes can be usefully parameterised in terms
of $\Delta^{\prime}\rho_{i}\hat{\beta}$. At low values one finds
the drift-tearing mode with frequency $\hat{\omega}\approx1+c\eta_{e},$
where $c$ varies weakly with $\eta_{e}$ (i.e. $c$ increases from
$0.7$ to $0.84$ as $\eta_{e}$ increases from small to large values),
has a growth rate $\hat{\gamma}\sim(\delta_{0}/\rho_{i})(\Delta^{\prime}-\Delta_{crit}^{\prime})$
if $\Delta^{\prime}>\Delta_{crit}^{\prime}\sim\rho_{i}^{-1}\hat{\beta}\eta_{e}^{2}(\ln(\rho_{i}/\delta_{0})+c^{\prime}\eta_{e}^{1/2}),$
similar to the stabilisation by large ion orbits shown in Ref. \citep{cowley:3230};
the term proportional to the constant $c^{\prime}$ resembles that
obtained for the cold ion model of Ref. \citep{drake:2509} {[}see
Eq. \eqref{eq:critdelprime} for a precise description]. As $\Delta^{\prime}\rho_{i}\hat{\beta}$
increases the drift-tearing mode is affected by a kinetic Alfvén wave
(KAW) propagating in the electron direction. Indeed for $\Delta^{\prime}\rho_{i}\hat{\beta}\sim\mathcal{O}(1)$
these couple strongly and there is a cross-over  with an exchange
of stability character: the mode identified by $\hat{\omega}\approx1+c\eta_{e}$
becomes stable, whereas it is the KAW that becomes unstable (see Fig.
\ref{fig:fig1}). Near the cross-over point the growth rate is enhanced:
$\hat{\gamma}\sim(\delta_{0}/\rho_{i})^{1/2}$; however, as $\Delta^{\prime}\rho_{i}\hat{\beta}$
increases further, the growth rate diminishes as $\hat{\gamma}\sim(\delta_{0}/\rho_{i})(\Delta^{\prime}\rho_{i}\hat{\beta}^{3})^{-1}$;
and $\hat{\omega}\rightarrow1$. Indeed, according to Eq. \eqref{eq:growthKAW},
when $\Delta^{\prime}$ exceeds a critical value, $\Delta_{crit}^{\prime}\sim\delta_{0}^{-1}\eta_{e}/\hat{\beta}$,
it becomes stable. Stability is also ensured if $\eta_{e}>9.4$.

We have also separately examined the stability of the $\hat{\omega}\approx1$
mode in the limit $\hat{\beta}\gg1$ and, somewhat surprisingly, find
results closely resembling those obtained in Ref. \citep{drake:2509}
for the cold ion limit: the mode
is completely stabilised for all $\Delta^{\prime}$ due to shielding
of the electron reconnecting layer as a result of the presence of
radial gradients in the electron temperature. To elucidate the transition
between instability at low $\hat{\beta}$ and stability at high $\hat{\beta}$,
we have developed a treatment for finite $\hat{\beta}$, but one which
is only strictly valid for a specific, though reasonable, value of
$\eta_{e}$ ($\eta_{e}=2.5$). This demonstrates the onset of stabilisation
due to plasma gradients when $\eta_{e}\hat{\beta}=0.85$. Below this critical value, $\hat\beta_c,$ the critical
$\Delta^{\prime}$ for stability is found to increase as $\hat{\beta}$
increases, as shown in Eq.\eqref{eq:last}, which is consistent with Eq.\eqref{eq:stabKAW}. A similar analytic calculation is also possible for the
much larger value $\eta_{e}=9.4,$ but in this case no unstable mode
is found, irrespective of the value of $\hat{\beta}.$ This is consistent
with the threshold $\eta_{e}=9.4$ predicted by low $\hat{\beta}$
theory. These stability results are summarised in Fig. \ref{fig:fig7}.

When the ideal internal kink mode stability parameter $\lambda_{H}=-\pi/\Delta^{\prime}r_{s}>0$,
our low $\hat{\beta}$ theory recovers results of Ref. \citep{pegoraro:364}
for the effect of the large ion orbits on the ideal internal kink
mode, although the role of the small electron dissipation is somewhat
different due to the improved electron model used here, which allows
electron temperature perturbations and hence involves the electron
temperature gradient parameter, $\eta_{e}$. When $\lambda_{H}=0$,
the dissipative internal kink mode, which is related to the solution
of our unified dispersion relation and whose stability depends entirely
on the electron effects, remains unstable. However, we find that this
mode becomes stable at a critical value of $\Delta^{\prime}$, namely
$\Delta^{\prime}r_{s}\equiv-\pi/\lambda_{H}=\delta_{0}^{-1}\hat{\beta}/\ln[(\rho_{i}\hat{\beta^{2}})/\delta_{0}]$.
Furthermore, if $\eta_{e}>2.38$ the mode is stable for all $\Delta^{\prime}.$

Finally we have exploited the low frequency of the dissipative internal
kink mode to develop an analytic dispersion relation, valid for finite
$\hat{\beta},$ to examine the effect of $\hat{\beta}$ on its stability.
An analytic expression \eqref{eq:kinkbounfarb} for the marginally
stable value can still be obtained. Although at first the critical
value of $-\lambda_{H}$ reduces as $\hat{\beta}$ is increased, this
trend eventually reverses and indeed all values of $-\lambda_{H}$
are unstable when $\hat{\beta}>7/18$. Figure \ref{fig:fig9} summarises
these results.

The combined stability diagram for the drift-tearing mode and the
dissipative internal kink mode shown in Fig. \ref{fig:fig11} provides
a basis for sawtooth modelling as described in Section \ref{sec:Sawtooth-Modelling}.
In the aftermath of a sawtooth crash one envisages a rapid recovery
of density and temperature profiles, followed by a slower evolution
of the $q$ profile. Essentially this leads to a decrease in the $\hat{\beta}$
parameter until the instability boundary for the dissipative internal
kink is encountered, triggering a further crash. This could be modelled
more completely with transport codes in the spirit of Ref. \citep{0741-3335-38-12-010},
but using the more precise stability criteria given in the present
theory. A simulation of the time evolution of the magnetic shear using
neoclassical resistivity and following its consequences for the trajectory
of $\hat{\beta}(t)$ in the stability diagram, leads to plausible
sawtooth periods for JET and ITER.

At this point it is helpful to highlight our key results, expressing
them in terms of the parameters $\hat{\beta}$ and $\delta_{0}/\rho_{i}\propto\tau^{1/2}C^{1/2}$;
the $\mathcal{O}(1)$ coefficients $c_{1}-c_{4}$ below are functions
of the other parameters, $\eta_{e},$ $\eta_{i}$, and $\tau$, although
we take $\tau=1$ in the evaluation of critical values of $\eta_{e}$
and \textbf{$\hat{\beta}$ }(more detailed descriptions of these appear
in the referenced equations):

\begin{itemize}
\item At low $\hat{\beta}$ the drift-tearing mode with $\hat{\omega}\approx1+0.8\eta_{e}$
is unstable for $\rho_{i}\Delta^{\prime}>c_{1}\hat{\beta}\eta_{e}^{2}\ln(\rho_{i}/\delta_{0})$
due to finite ion orbit effects {[}see Eq. \eqref{eq:critdelprime}].
\item At $\Delta^{\prime}\rho_{i}\hat{\beta}\sim\mathcal{O}(1)$ this mode
interacts with the KAW and for larger values of this parameter it
is the KAW with $\hat{\omega}\approx1$ that is unstable, until it
is stabilised when $\Delta^{\prime}>c_{2}\hat{\beta}/\delta_{0}$
{[}see Eq. \eqref{eq:stabKAW}]; it is also stable for all $\Delta^{\prime}$
if $\eta_{e}>9.4.$
\item For high $\hat{\beta}$ there is stability of the tearing mode for
all $\Delta^{\prime}$, the transition occurring when $\hat{\beta}=c_{3}/\eta_{e}$
{[}see Eq. \eqref{eq:stabmia}]; the critical $\Delta^{\prime}$ in
Eq. \eqref{eq:stabKAW} also increases as $\hat{\beta}$ approaches this
transition value, as shown in Eq. \eqref{eq:ecche}.
\item The stability regimes of the drift-tearing and kinetic Alfv\'en
waves in terms of $\hat\beta$ and $\Delta^{\prime}\rho_i$ are summarised in Fig. \ref{fig:fig7}.
\item At low $\hat{\beta}$ the dissipative internal kink mode is only unstable
for $\Delta^{\prime}\rho_{i}\equiv-\pi\rho_{i}/(r_{s}\lambda_{H})>c_{4}\hat{\beta}/[\delta_{0}/\rho_{i}\ln(\hat{\beta}^{2}\rho_{i}/\delta_{0})]$
{[}see Eq. \eqref{eq:critlambdaH}]; it is also stable for all $\Delta^{\prime}$
if $\eta_{e}>2.4.$
\item At higher $\hat{\beta}>7/18$, according to Eq. \eqref{eq:kinkbounfarb}
there is instability for all $-\lambda_{H}$.
\item Figures \ref{fig:fig9} and \ref{fig:fig10} provide stability diagrams
for the drift-tearing/KAW and dissipative internal kink modes in the
space of $\Delta^{\prime}$(or $-\lambda_{H}$) and $\hat{\beta}$ for higher values of $\Delta^{\prime}\rho_i.$ 
\item Figure \ref{fig:fig11} suggests a resulting scenario for the onset
of a sawtooth crash .
\end{itemize}
Our calculation is limited to cylindrical geometry \textendash{} strictly
to a sheared slab since we neglect effects of cylindrical curvature.
In toroidal geometry one must take account of magnetic drifts. In
the limit that the ion banana width, $w_{ban},$ is small compared
to $\delta_{0},$ one can derive similar equations to those of Ref.
\citep{drake:2509}, but encompassing neoclassical transport effects
\citep{10.1063/1.859173,10.1063/1.859387}. While we may neglect most
of these for small values of the magnetic shear and the fraction of
trapped particles, an important consequence arises from the ion neoclassical
polarisation drift which reduces the parameter $C$ by a factor $B_{p}^{2}/B_{T}^{2}\ll1,$
where $B_{p}$ and $B_{T}$ are the poloidal and toroidal magnetic
fields. In the opposite limit $w_{ban}\gg\delta_{0}$ one would need
to include the complete drift orbits; this calculation would be extremely
complicated but can be expected to be qualitatively similar to the
large ion Larmor orbit case with the substitution $\rho_{i}\rightarrow w_{ban}$.
The fact that the effective value of $C$ is greatly reduced in the
torus significantly extends the experimental validity of our large
orbit theory.
\begin{acknowledgments}
Part of this work was carried out at a Workshop on Gyro-kinetics held
under the auspices of the Isaac Newton Institute for Mathematical
Sciences, Cambridge, in July and August 2010. This work was jointly
funded by the United Kingdom Physical Science and Engineering Research
Council and Euratom. A.Z. was supported by the Leverhulme Trust Network
for Magnetized Plasma Turbulence, a Culham Fusion Research Fellowship
and an EFDA Fellowship. The views and opinions expressed herein do
not necessarily reflect those of the European Commission.
\end{acknowledgments}

\appendix

\section{Internal Kink Mode}

To address the case of finite $\hat{\beta}$ for the dissipative internal
kink mode, we can consider the limit $|\hat{\omega}|\ll1$, unlike
the situation with the $m>1$ tearing mode which has $\omega\sim\omega_{*e}$.
In the electron layer, region $2$, Eq. \eqref{eq:electr} then becomes\begin{equation}
\frac{d^{2}A}{ds^{2}}=\hat{\omega}\hat{\beta}\sigma(s)A\label{eq:app1}\end{equation}
where\begin{equation}
\sigma(s)=\frac{1+1.71\eta_{e}+d_{1}s^{2}}{1+[d_{0}-(1+1.71\eta_{e})]s^{2}+(1+\tau)\hat{\omega}s^{4}}.\label{eq:app2}\end{equation}
In the region $s\sim 1$ we can ignore the last term in the denominator
and solve Eq. \eqref{eq:app1} as an expansion in $\hat{\omega},$
retaining it only for large $s\sim|\hat{\omega}|^{-1/2}$ , when it
competes with the term proportional to $s^{2}$. Thus, setting $A=A_{0}$
in lowest order we find the solution even at $s=$0 to be:\begin{equation}
A=A_{0}\left[1+\hat{\beta}\hat{\omega}\int_{0}^{s}ds^{\prime}(s-s^{\prime})\sigma(s^{\prime})\right].\end{equation}
At large $s$ we find\begin{equation}
A\sim A_{0}\left[1+\frac{\pi}{2}\sqrt{d}\left(\frac{a}{d}-\frac{d_{1}}{d^{2}}\right)\hat{\beta}\hat{\omega}s\right]\,\,\,\mbox{for}\,\,\, s\rightarrow\infty,\label{eq:app3}\end{equation}
with $a=1+1.71\eta_{e},$ $d=d_{0}-a.$ In the region $s\sim|\hat{\omega}|^{1/2},$
Eq. \eqref{eq:app1} takes the form\begin{equation}
\frac{d^{2}A}{dy^{2}}=\frac{\hat{\beta}}{1+\tau}\frac{A}{1+y^{2}},\end{equation}
where $y^{2}=d^{-1}d_{1}(1+\tau)\hat{\omega}s^{2},$ with solution\begin{equation}
A=b_{1}\,_{2}F_{1}\left(-\frac{1}{4}-\frac{\mu_{1}}{2},-\frac{1}{4}+\frac{\mu_{1}}{2};\frac{1}{2};-y^{2}\right)+b_{2}\,_{2}F_{1}\left(\frac{1}{4}-\frac{\mu_{1}}{2},\frac{1}{4}+\frac{\mu_{1}}{2};\frac{3}{2};-y^{2}\right),\label{eq:app4}\end{equation}
where $\,_{2}F_{1}$ is the hypergeometric function \citep{abramhyper}
and $\mu_{1}=\sqrt{1/4+\hat{\beta}/(1+\tau)}.$ Matching to solution
\eqref{eq:app3} at small $y$ we have\begin{equation}
\frac{b_{2}}{b_{1}}=\frac{\pi}{2}\hat{\beta}\left(a-\frac{d_{1}}{d}\right)\sqrt{\frac{\hat{\omega}}{d_{1}(1+\tau)}}.\end{equation}
To match to the ion region $1,$ we need the large $y$ limit of solution
\eqref{eq:app4} \begin{equation}
\begin{split} & A\sim y^{1/2+\mu_{1}}\left[1+\frac{b_{2}}{2b_{1}}\frac{\Gamma\left(\frac{3}{4}+\frac{\mu_{1}}{2}\right)}{\Gamma\left(\frac{5}{4}+\frac{\mu_{1}}{2}\right)}\frac{\Gamma\left(-\frac{1}{4}+\frac{\mu_{1}}{2}\right)}{\Gamma\left(\frac{1}{4}+\frac{\mu_{1}}{2}\right)}\right]\\
 & +y^{1/2-\mu_{1}}\frac{\Gamma(-\mu_{1})}{\Gamma(\mu_{1})}\frac{\Gamma\left(\frac{3}{4}+\frac{\mu_{1}}{2}\right)}{\Gamma\left(\frac{3}{4}-\frac{\mu_{1}}{2}\right)}\frac{\Gamma\left(-\frac{1}{4}+\frac{\mu_{1}}{2}\right)}{\Gamma\left(-\frac{1}{4}-\frac{\mu_{1}}{2}\right)}\left[1+\frac{b_{2}}{2b_{1}}\frac{\Gamma\left(\frac{3}{4}-\frac{\mu_{1}}{2}\right)}{\Gamma\left(\frac{1}{4}-\frac{\mu_{1}}{2}\right)}\frac{\Gamma\left(-\frac{1}{4}-\frac{\mu_{1}}{2}\right)}{\Gamma\left(\frac{5}{4}-\frac{\mu_{1}}{2}\right)}\right].\end{split}
\end{equation}
Since in the ion region we solve for the Fourier transformed current
$\hat{J}(k),$ the form of $J(s)$ in the electron region can be obtained
using Amp\'ere's law. Thus, we obtain\begin{equation}
\frac{b_{-}}{b_{+}}=\left[\frac{d}{d_{1}(1+\tau)\hat{\omega}}\right]^{\mu_{1}}\frac{1/2+\mu_{1}}{1/2-\mu_{1}}\frac{\Gamma(-\mu_{1})}{\Gamma(\mu_{1})}\frac{\Gamma^{2}\left(\frac{3}{4}+\frac{\mu_{1}}{2}\right)}{\Gamma^{2}\left(\frac{3}{4}-\frac{\mu_{1}}{2}\right)}\frac{\Xi+\tan\left[\frac{\pi}{2}\left(\frac{1}{2}-\mu_{1}\right)\right]}{\Xi+\tan\left[\frac{\pi}{2}\left(\frac{1}{2}+\mu_{1}\right)\right]},\label{eq:app5}\end{equation}
where\begin{equation}
\Xi=\frac{1}{\pi(a-d_{1}/d)}\sqrt{\frac{d_{1}}{(1+\tau)\hat{\omega}}}\frac{\Gamma\left(\frac{1}{4}-\frac{\mu_{1}}{2}\right)}{\Gamma\left(\frac{3}{4}-\frac{\mu_{1}}{2}\right)}\frac{\Gamma\left(\frac{1}{4}+\frac{\mu_{1}}{2}\right)}{\Gamma\left(\frac{3}{4}+\frac{\mu_{1}}{2}\right)}.\label{eq:app13}\end{equation}
Noting that $|\Xi|\sim|\hat{\omega}|^{-1/2}\gg1$ when $|\hat{\omega}|\ll1,$
in Eq. \eqref{eq:app5}, we find that the low $\hat{\beta}$ limit
of \eqref{eq:app5} coincides with the low $\hat{\beta}$ calculation
\eqref{eq:elsollow} at low $\hat{\omega}.$

Equation \eqref{eq:app5}, after using Eq. \eqref{eq:rule}, must be matched
to $\hat{a}_{-}/\hat{a}_{+}$ as derived from the region $1$ solution.
An analytic expression for this ratio can be obtained by iterating
the solution in $|\hat{\omega}|^{2}\hat{\beta}\ll1,$ in a similar manner
to our low $\hat{\beta}$ treatment of the drift-tearing mode. Thus,
in place of Eq. \eqref{eq:ion} we have\begin{equation}
\frac{d}{dk}\left[\frac{G_{0}(k)}{F_{0}(k)}\frac{d\hat{J}}{dk}\right]=\hat{\omega}\hat{\beta}\frac{\hat{J}}{k^{2}},\label{eq:app9}\end{equation}
where 

\begin{equation}
G_{0}(k)=-(1-\hat{\omega}^{-1})+F_{0}(k),\label{eq:app7}\end{equation}
\begin{equation}
F_{0}(k)=\hat{\omega}^{-1}\left\{ \Gamma_{0}(k^{2}/2)-1-\eta_{i}k^{2}/2\left[\Gamma_{0}(k^{2}/2)-\Gamma_{1}(k^{2}/2)\right]\right\} \label{eq:app8}\end{equation}
where we retain the $\mathcal{O}(\hat{\omega})$ correction in $G_{0}$
for large $k,$ since in this limit $F_{0}\sim k^{-1}.$ Solution
\eqref{eq:pegsolions} is then modified to give\begin{equation}
\begin{split} & \hat{J}(k)=\exp\left[-\hat{\omega}\hat{\beta}\int_{0}^{k}\frac{du}{u}\frac{F_{0}(u)}{G_{0}(u)}\right]+\hat{\omega}\hat{\beta}\frac{\lambda_{H}}{\rho_{i}}\int_{0}^{k}du\frac{F_{0}(u)}{G_{0}(u)}\\
 & -\left(\hat{\omega}\hat{\beta}\right)^{2}\int_{0}^{k}du\frac{F_{0}(u)}{G_{0}(u)}\int_{0}^{u}dv\frac{F_{0}(v)}{v^{2}G_{0}(v)},\end{split}
\label{eq:app6}\end{equation}
where we have expressed $\Delta^{\prime}$ in terms of $\lambda_{H}.$
To match solution \eqref{eq:app6} to that in region $2,$ we must
first consider an intermediate large $k$ range where the approximations
\eqref{eq:app7}-\eqref{eq:app8} fail: $k\sim|\hat{\omega}|^{-1}\gg1.$
In this region \begin{equation}
\frac{G_{0}}{F_{0}}=-\hat{\omega}(1+\tau)+\frac{1}{\sqrt{\pi}}\left(1-\frac{\eta_{i}}{2}\right)\frac{1}{k},\label{eq:app12}\end{equation}
so that Eq. \eqref{eq:app9} can be written as\begin{equation}
u(1-u)\frac{d^{2}\hat{J}}{du^{2}}-\frac{d\hat{J}}{du}+\frac{\hat{\beta}}{1+\tau}\hat{J}=0\end{equation}
where $u=\sqrt{\pi}\hat{\omega}(1+\tau)(1-\eta_{i}/2)^{-1}k,$ with
solution \begin{equation}
\hat{J}(k)=a_{1}\,_{2}F_{1}\left(-\frac{1}{2}-\mu_{1},-\frac{1}{2}+\mu_{1};1;1-u\right)+a_{2}u^{2}\,_{2}F_{1}\left(\frac{3}{2}-\mu_{1},\frac{3}{2}+\mu_{1};3;u\right).\label{eq:app10}\end{equation}
{[}Note, the second solution is a special case, see p. 75, Eq. (7)
of Ref.\citep{erdelyi2}]. For $|\hat{\omega}k|\ll1,$ i.e. $u\ll1,$
solution \eqref{eq:app10} takes the form\begin{equation}
\begin{split} \hat{J}(k)&\sim a_{1}\left\{ \frac{1-(1/4-\mu^{2})u}{\Gamma\left(\frac{3}{2}+\mu_{1}\right)\Gamma\left(\frac{3}{2}-\mu_{1}\right)}+\frac{u^{2}(k_{0}-\ln u)}{2\Gamma\left(-\frac{1}{2}+\mu_{1}\right)\Gamma\left(-\frac{1}{2}-\mu_{1}\right)}+\ldots\right\} \\
& +a_{2}u^{2}(1+\ldots)\,\,\,\mbox{for\,\,\,}|\hat{\omega}k|\ll1,\end{split}
\label{eq:app11}\end{equation}
with $k_{0}=\psi(1)+\psi(3)-\psi(3/2-\mu_{1})-\psi(3/2+\mu_{1}),$
where $\psi$ is the digamma function \citep{abramgamma}. Expression \eqref{eq:app11}
can be matched to the large $k$ limit of solution \eqref{eq:app6}
gwhere the form \eqref{eq:app12} implies \begin{equation}
\frac{F_{0}}{G_{0}}\approx\frac{\sqrt{\pi}}{1-\eta_{i}/2}k\left[1+\frac{\sqrt{\pi}\hat{\omega}(1+\tau)}{1-\eta_{i}/2}k\right],\end{equation}
so that\begin{equation}
\begin{split} & \hat{J}(k)=1+\frac{\sqrt{\pi}\hat{\omega}\hat{\beta}}{1-\eta_{i}/2}k+\frac{\pi}{2}\left(\frac{\hat{\omega}k}{1-\eta_{i}/2}\right)^{2}\hat{\beta}(\hat{\beta}+1+\tau)\\
 & -\frac{\pi}{2}\left(\frac{\hat{\omega}k\hat{\beta}}{1-\eta_{i}/2}\right)^{2}\left[\left(1-\frac{\eta_{i}}{2}\right)I_{2}+\ln k-\frac{1}{2}\right]-\frac{\sqrt{\pi}\hat{\omega}\hat{\beta}}{2(1-\eta_{i}/2)}\frac{\lambda_{H}}{\rho_{i}}k^{2},\end{split}
\end{equation}
where\begin{equation}
I_{2}=\frac{1}{\sqrt{\pi}}\int_{0}^{\infty}\frac{dk}{k^{2}}\left[\frac{F_{0}}{G_{0}}-\frac{\sqrt{\pi}}{1-\eta_{i}/2}\frac{k^{2}}{1+k}\right]\end{equation}
(in this integral we can take the $\hat{\omega}\rightarrow0$ limit
of $F$ anf $G$). The result of the matching is\begin{equation}
\begin{split} & \frac{a_{2}}{a_{1}}=-\frac{\cos(\pi\mu_{1})}{2\pi}\left\{ 1-\frac{1}{\sqrt{\pi}}\frac{1-\eta_{i}/2}{1+\tau}\frac{\lambda_{H}}{\rho_{i}}\frac{1}{\hat{\omega}}-\hat{\beta}\frac{1-\eta_{i}/2}{1+\tau}I_{2}\right.\\
 & \left.+\frac{\hat{\beta}}{1+\tau}\left[\frac{3}{2}-k_{0}+\ln\left(\frac{\sqrt{\pi}(1+\tau)}{1-\eta_{i}/2}\hat{\omega}\right)\right]\right\} .\end{split}
\label{eq:app22}\end{equation}
It remains to match the large $u$ limit of solution \eqref{eq:app10}\begin{equation}
\begin{split} & \hat{J}(k)\sim\frac{\Gamma(2\mu_{1})e^{i\pi\mu_{1}}}{\Gamma\left(\frac{3}{2}+\mu_{1}\right)}\left(2\frac{a_{2}}{a_{1}}+ie^{i\pi\mu_{1}}\frac{\hat{\beta}}{1+\tau}\right)u^{1/2+\mu_{1}}\\
 & +\frac{\Gamma(-2\mu_{1})e^{-i\pi\mu_{1}}}{\Gamma\left(\frac{3}{2}-\mu_{1}\right)}\left(2\frac{a_{2}}{a_{1}}+ie^{-i\pi\mu_{1}}\frac{\hat{\beta}}{1+\tau}\right)u^{1/2-\mu_{1}},\end{split}
\label{eq:app23}\end{equation}
to the region $2$ solution. Reintroducing $k$, Eq. \eqref{eq:app23}
provides the ratio $\hat{a}_{2}/\hat{a}_{1}$ to be matched to $\hat{c}_{+}/\hat{c}_{-}$
as given by Eqs. \eqref{eq:app5} with the help of Eq. \eqref{eq:rule},
i.e. \begin{equation}
\frac{\hat{c}_{-}}{\hat{c}_{+}}=\frac{\Gamma\left(-\frac{1}{2}+\mu_{1}\right)}{\Gamma\left(-\frac{1}{2}-\mu_{1}\right)}\tan\left[\frac{\pi}{2}\left(\frac{1}{2}+\mu_{1}\right)\right]\frac{b_{+}}{b_{-}}.\end{equation}
This yields the final dispersion relation\begin{equation}
\begin{split}\left[\frac{e^{-i\frac{5}{4}\pi}}{8\sqrt{\pi}\hat{\omega}}\frac{\delta_{0}}{\rho_{i}}\sqrt{\frac{d(\eta_{e})}{d_{1}}}\,\frac{1-\eta_{i}/2}{(1+\tau)^{3/2}}\right]^{2\mu_{1}}\times\frac{\Xi(\hat{\omega})+\tan\left[\frac{\pi}{2}\left(\frac{1}{2}-\mu_{1}\right)\right]}{\Xi(\hat{\omega})+\tan\left[\frac{\pi}{2}\left(\frac{1}{2}+\mu_{1}\right)\right]}= & \\
 \frac{1/2-\mu_{1}}{1/2+\mu_{1}}\frac{\Gamma^{2}(\mu_{1})}{\Gamma^{2}(-\mu_{1})}\frac{\Sigma(\hat{\omega})+i\frac{\hat{\beta}}{1+\tau}e^{i\pi\mu_{1}}}{\Sigma(\hat{\omega})+i\frac{\hat{\beta}}{1+\tau}e^{-i\pi\mu_{1}}},&\end{split}
\label{eq:dispkinkarbbetaappwhole}\end{equation}
where $\Sigma(\hat{\omega})=2a_{2}/a_{1}.$ We note that we can simplify
Eq. \eqref{eq:dispkinkarbbetaappwhole} since $|\Xi(\hat{\omega})|\propto\left|\hat{\omega}\right|^{-1/2}\gg1,$
as given by Eq. \eqref{eq:app13},\begin{equation}
\left[\frac{e^{-i\frac{5}{4}\pi}}{8\sqrt{\pi}\hat{\omega}}\frac{\delta_{0}}{\rho_{i}}\sqrt{\frac{d(\eta_{e})}{d_{1}}}\,\frac{1-\eta_{i}/2}{(1+\tau)^{3/2}}\right]^{2\mu_{1}}=\frac{1/2-\mu_{1}}{1/2+\mu_{1}}\frac{\Gamma^{2}(\mu_{1})}{\Gamma^{2}(-\mu_{1})}\frac{\Sigma(\hat{\omega})+i\frac{\hat{\beta}}{1+\tau}e^{i\pi\mu_{1}}}{\Sigma(\hat{\omega})+i\frac{\hat{\beta}}{1+\tau}e^{-i\pi\mu_{1}}}.\label{eq:dispkinkarbbetaapp}\end{equation}
Equation \eqref{eq:dispkinkarbbetaapp} provides a finite $\hat{\beta}$
dispersion relation for the dissipative internal kink mode, where
we recall $\delta_{0}/\rho_{i}\ll1$ and $\mu_{1}$ is independent
of $\hat{\omega}.$

\vspace{10cm}


\end{document}